%% file: paper.tex
\documentstyle[11pt,aaspp4]{article}

\def\arcs{\ifmmode {^{\scriptscriptstyle\prime\prime}}
          \else $^{\scriptscriptstyle\prime\prime}$\fi}
\def\arcm{\ifmmode {^{\scriptscriptstyle\prime}}
          \else $^{\scriptscriptstyle\prime}$\fi}
\newdimen\sa  \newdimen\sb
\def\parcs{\sa=.07em \sb=.03em
     \ifmmode $\rlap{.}$^{\scriptscriptstyle\prime\kern -\sb\prime}$\kern -\sa$
     \else \rlap{.}$^{\scriptscriptstyle\prime\kern -\sb\prime}$\kern -\sa\fi}
\def\parcm{\sa=.08em \sb=.03em
     \ifmmode $\rlap{.}\kern\sa$^{\scriptscriptstyle\prime}$\kern-\sb$
     \else \rlap{.}\kern\sa$^{\scriptscriptstyle\prime}$\kern-\sb\fi}
\def\pdeg{\ifmmode $\setbox0=\hbox{$^{\circ}$}\rlap{\hskip.11\wd0 .}$^{\circ}
          \else \setbox0=\hbox{$^{\circ}$}\rlap{\hskip.11\wd0 .}$^{\circ}$\fi}
\def\gtorder{\mathrel{\raise.3ex\hbox{$>$}\mkern-14mu
             \lower0.6ex\hbox{$\sim$}}}
\def\ltorder{\mathrel{\raise.3ex\hbox{$<$}\mkern-14mu
             \lower0.6ex\hbox{$\sim$}}}

\input psfig
\lefthead{Falco et al.}
\righthead{}

\begin{document}

\title{Dust and Extinction Curves in Galaxies with z$>$0: \\
   The Interstellar Medium of Gravitational Lens Galaxies\footnote{Based on 
Observations made with the NASA/ESA Hubble Space
Telescope, obtained at the Space Telescope Science Institute, which is
operated by AURA, Inc., under NASA contract NAS 5-26555. }} 

\vskip 2truecm

\author{E.E. Falco$^{(a)}$,} 
\author{C.D. Impey$^{(b)}$, C.S. Kochanek$^{(a)}$, J. Leh\'ar$^{(a)}$, B.A. McLeod$^{(a)}$}
\author{H.-W. Rix$^{(b)}$, C.R. Keeton$^{(b)}$, J.A. Mu\~noz$^{(a)}$ and C.Y. Peng$^{(b)}$}

\affil{$^{(a)}$ Harvard-Smithsonian Center for Astrophysics, 60 Garden
	St., Cambridge, MA 02138}
\affil{email: ckochanek, efalco, jlehar, bmcleod, jmunoz@cfa.harvard.edu}
\affil{$^{(b)}$Steward Observatory, University of Arizona, Tucson, AZ 85721}
\affil{email: impey, rix, ckeeton, cyp@as.arizona.edu}

\begin{abstract}
We determine 37 differential extinctions in 23 gravitational lens
galaxies over the range $0 \ltorder z_l \ltorder 1$.  Only 7 of the 23
systems have spectral differences consistent with no differential
extinction. The median differential extinction for the
optically-selected (radio-selected) subsample is $\Delta E(B-V)=0.04$
($0.06$) mag.  The extinction is patchy and shows no correlation with
impact parameter.  The median total extinction of the bluest images is
$E(B-V)=0.08$ mag, although the total extinction distribution is
dominated by the uncertainties in the intrinsic colors of quasars.
The directly measured extinction distributions are consistent with the
mean extinction estimated by comparing the statistics of quasar and
radio lens surveys, thereby confirming the need for extinction
corrections when using the statistics of lensed quasars to estimate
the cosmological model.  A disjoint subsample of two face-on,
radio-selected spiral lenses shows both high differential and total
extinctions, but standard dust-to-gas ratios combined with the
observed molecular gas column densities overpredict the amount of
extinction by factors of 2--5. For several systems we can estimate the
extinction law, ranging from $R_V=1.5\pm0.2$ for a $z_l=0.96$
elliptical, to $R_V=7.2\pm0.1$ for a $z_l=0.68$ spiral.  For the four
radio lenses where we can construct non-parametric extinction curves
we find no evidence for gray dust over the IR--UV wavelength range.
The dust can be used to estimate lens redshifts with reasonable
accuracy, although we sometimes find two degenerate redshift
solutions.
\end{abstract}


\section{Introduction}

Corrections for extinction have significant cosmological consequences
for determining the Hubble constant (e.g. Freedman et al. 1998), the
cosmological model using either Type Ia supernovae (Perlmutter et
al. 1997, Riess et al. 1998) or gravitational lenses (Kochanek 1996,
Falco, Kochanek \& Mu\~noz 1998), and the epoch of star formation
(e.g. Madau, Pozetti \& Dickinson 1998). Yet detailed, quantitative
studies of extinction are largely limited to galaxies closer than
$\ltorder 10$~Mpc where it is possible to study individual stars.
Studies of extinction at larger distances rely on analyzing the
spectral energy distributions of stars mixed with dust, and the amount
of dust inferred for a given change in color depends strongly on
geometric and other assumptions (e.g. Witt, Thronson \& Capuano 1992).
The patchy mixture of stars, dust and star forming regions in nearby
galaxies illustrates the complexities.

The extinction curves which determine the relation between
changes in color and total absorption are reliably measured only in
the Galaxy, the LMC and the SMC (see Mathis 1990, Fitzpatrick 1998).
We will describe extinction laws by $R_\lambda$, where the magnitude
change at wavelength $\lambda$ for extinction $E(B-V)$ is $A_\lambda =
R_\lambda E(B-V)$.  Most, but not all (Mathis \& Cardelli 1992), of
the extinction curve variations seen in these three galaxies can be
described by parametric functions of the value in the V band, $R_V$
(e.g. Savage \& Mathis 1979, Fitzpatrick \& Massa 1988, Cardelli et
al. 1989, Fitzpatrick 1998).  The variations tend to be associated
with colder, denser regions of the interstellar medium, frequently in
directions with appreciable molecular gas (e.g.  Jenniskens \&
Grennberg 1993).  The variations affect mainly the optical and
ultraviolet extinction curve. The infrared and near-infrared
extinction curve is a fixed power law with $ R_\lambda \propto
\lambda^{-\alpha}$ and $\alpha \simeq 1.7 \pm 0.1$ (Mathis 1990).  In
the Galaxy, typical paths with modest total extinctions have
$R_V\simeq 3.1$, while paths through denser, higher extinction regions
can reach an $R_V > 5$ for an overall range of $2.2 \ltorder R_V
\ltorder 5.8$.  The changes in the optical extinction curve are
correlated with the width and amplitude of the 2175\AA\ PAH feature.
In the SMC and a few regions of the LMC, however, the ultraviolet
extinction curve differs significantly from the standard Galactic law
with the same optical extinction properties.  The principal difference
is that these regions have a far weaker 2175\AA\ feature (e.g. Gordon
\& Clayton 1998).  Physically, the extinction law depends on the mean
size and composition distribution of the dust grains along the line of
sight (e.g. Draine \& Malhotra 1993, Rouleau, Henning \& Stognienko
1997), so it is not surprising that the properties of the extinction
curve depend on the environment.

The Galaxy, the LMC, and the SMC do not constitute a representative
sample of galaxy types, ages, metallicities, or star formation
histories, and the available data on extinction curves in other
galaxies provide little evidence that the typical extinction law
observed in the Galaxy is also typical for other galaxies.  Estimates
in M31 produce a range of $3.5$ to $5.3$ for $R_V$ where the
variations appear to be linked to the local metallicity (Walterbos
1986, Hodge \& Kennicutt 1982, Iye \& Richter 1985).  There is some
evidence that extinction curves in early-type galaxies are
significantly steeper as a function of $\lambda^{-1}$, with $1.9
\ltorder R_V \ltorder 2.8$ (e.g. Warren-Smith \& Berry 1983, Rifatto
1990, Brosch \& Loinger 1991, Goudfrooij et al. 1994).  Lower $R_V$
values correspond to a smaller average size for the dust grains
compared to the Galaxy.  Riess, Press \& Kirshner (1996) estimated a
mean extinction curve $R_V=2.55\pm0.30$ for 20 nearby Type Ia
supernovae with I, R, V and B band photometry, which is marginally
consistent with the typical Galactic value.  Despite the observed
variations in the extinction curve, the typical Galactic or extreme
SMC extinction curves are almost universally adopted for all other
galaxies from $z=0$ to $z>5$.  Extinction curves almost certainly
evolve with redshift since the metallicity, elemental abundance
ratios, mean star formation rate, and energy injection rates which
determine the structure and evolution of the dust are all strong
functions of redshift.

Recently, two new techniques for making direct measurements of dust in
high redshift galaxies have appeared.  The first new method is simply
an extension of local methods for stars to the far brighter Type Ia
supernovae (Riess et al. 1998, Perlmutter et al. 1997).  Accurate
extinction measurements and corrections are essential to the goal of
using the Type Ia supernovae to determine the cosmological model.  The
second new method is to use multiply imaged gravitational lenses.  As
first pointed out by Nadeau et al. (1991), the lenses can be used to
determine the differential extinction between the images and the
extinction law in the lens galaxy.  Jean \& Surdej (1998) noted that
it is also possible to determine the redshift of the dust in systems
with significant differential extinction.  Malhotra, Rhoads \& Turner
(1997) attempted the first survey of extinction in lenses based on the
overall optical and infrared colors of the systems, concluding that
many lens galaxies had large, uniform dust opacities.  However, high
resolution imaging has now shown that many of the radio lenses with
red broad band colors are red because the flux from the source is
dominated by light from the AGN host galaxy (see Kochanek et
al. 1998b).

The Center for Astrophysics/University of Arizona Space Telescope Lens
Survey (CASTLES) is systematically obtaining V, I and H band
photometry of the nearly $50$ known lens systems.  The broad
wavelength coverage makes accurate determinations of the extinction in
point-source lens systems relatively simple.  The substantial number
of lenses with extended optical and infrared sources are more
difficult to use; we will ignore them for the current survey.  By
combining our CASTLES data with archival HST and ground-based
photometry we can perform a preliminary survey of extinction in 23
gravitational lenses, 8 of which were radio-selected.  Unlike the
optically-selected systems, the radio-selected systems should be
unbiased with respect to the dust content of the lens galaxy.  In \S2
we describe the method for determining extinctions and extinction
laws.  In \S3 and \S4 we determine the differential and total
extinctions for the systems assuming a standard $R_V=3.1$ Galactic
extinction law.  In \S5 we estimate extinction laws and dust
redshifts.  In \S6 we discuss the implications of our results for
determining the cosmological model using gravitational lens statistics
and Type Ia supernovae.  We summarize our results and discuss future
expansions of the method in \S7.

\section{Methods}

For a gravitational lens, we can divide the study of extinction into
determining the differential extinction between the lines of sight
traversed by the images through the galaxy, and the total extinction
along the lines of sight.  The differential extinction is the more
easily and accurately measured because it requires only the assumption
that the spectrum of the source is identical for each of the images.
The total extinction is more difficult to determine because
it depends on a model for the intrinsic spectrum of the source and
combines extinction from the Galaxy, the lens, and the host.

\def\lsrc{{\lambda \over 1 + z_s}} 
\def\llens{{\lambda \over 1 + z_l}}
We observe $N_i$ lensed images of a point-source whose intrinsic
spectrum expressed in magnitudes at an observed wavelength $\lambda$
and time $t$ is $m_0(\lambda,t)$.  The images we see at time $t$ are
magnified by $M_i(\lambda,t)$ and have time delays $\Delta t_i$, so
that the observed spectrum of image $i$ at time $t$ is

\begin{equation}
 m_i(\lambda,t) = m_0(\lsrc,t-\Delta t_i) - 2.5 \log M_i(\llens,t) +
   E_i R_i(\llens) + E_{Gal} R_{Gal} (\lambda) + E_{src} R_{src}
   (\lsrc)
\end{equation}
where $E_i$, $E_{Gal}$ and $E_{src}$ are the extinctions $E(B-V)$ due to
the lens for image $i$, the Galaxy respectively, and the source's host galaxy, and
$R_i(\lambda)$, $R_{Gal}(\lambda)$, and $R_{src}(\lambda)$ are the
extinction laws for the lens at image $i$, the Galaxy, and the host
galaxy respectively.  The lens and host extinction curves must be redshifted by the
lens and host redshifts $z_l$ and $z_s$.  The images pass through the
lens at locations separated by kpc, so we expect different extinctions
and possibly even different extinction laws for each of the images.
The ray separations in the host galaxy or our Galaxy are so small
($\sim 10^{-5} d \ll 1$ pc for dust at distance $d < 1$ kpc from the source 
or observer) that we can safely assume that the host and the Galaxy contribute equal
extinctions to all images.  If (1) the magnification is wavelength
independent; (2) the magnification is time independent; (3) the source
spectrum is time independent for the both the duration of the
observations and the time delays; and (4) the extinction curves for
the images are identical, then the magnitude differences between
images $i$ and $j$,

\begin{equation}
 m_i(\lambda)-m_j(\lambda) = - 2.5 \log { M_i \over M_j} + (E_i-E_j)
 R \left( { \lambda \over 1+z_l }\right),
\end{equation}
depend only on the constant magnification ratios $M_i/M_j$, the
extinction differences $E_i-E_j$, and the extinction curve
$R(\lambda/(1+z_l))$ in the lens rest frame.  Nadeau et al.  (1991)
were the first to explore the dependence of the flux ratios on the
extinction and the extinction curve, while Jean \& Surdej (1998)
pointed out the potential utility of the dependence on the lens
redshift.  For a lens with $N_{im}=2$ or $4$ images with flux ratios
measured at $N_\lambda$ wavelengths we have $(N_{im}-1)N_\lambda$
constraints to determine $N_{im}-1$ flux ratios and differential
extinctions when the extinction law is fixed.  Two wavelengths are
sufficient for a measurement, but three are needed to check the
consistency of the solution.  For an extinction curve parameterized by
$R_V$, we need measurements at three or more wavelengths to estimate
$R_V$.  We can also determine the extinction curve non-parametrically
(following Nadeau et al. 1991).  For a two-image lens we can determine
the curve at $N_\lambda-2$ wavelengths, but we will always find a
perfect fit to the data and there will be no degrees of freedom left
to check the solution.  Four-image systems are better because the
multiple lines-of-sight add considerable redundancy.  Radio data are
particularly valuable because we can be certain that they are
completely unaffected by dust, allowing us to detect, limit or rule
out dust which is gray over the entire IR--UV wavelength range.

We estimate the differential extinctions by fitting the measured 
magnitude differences assuming no time dependence to the spectrum, wavelength
dependence for the magnification, or position dependence to the
extinction law.  Thus, we minimize

\begin{equation}
    \chi^2 = \sum_{j=1}^{N_\lambda} \sum_{i=1}^{N_{im}} 
     { \left(  m_i(\lambda_j) - m_0(\lambda_j) - 2.5 \log M_i - E_i R(\lambda_j/(1+z_l)) \right)^2 
              \over \sigma_{ij}^2 + \sigma_{sys}^2 } 
\end{equation}
over the $i=1 \cdots N_{im}$ images and $j=1 \cdots N_\lambda$ filters
for magnitude measurements $m_i(\lambda_j)$ with uncertainties
$\sigma_{ij}$.  We include an additional systematic error term
$\sigma_{sys}$ whose value is set to zero for the actual fits.  We
simply used the standard central wavelengths of the filters in the
calculations.  We set one magnification to unity and one extinction to
zero because we can only measure relative magnifications and
differential extinctions.  We always picked the bluest
(i.e. lowest-extinction) image to be the reference for the extinction.

The general expression in eqn. (1) illustrates the possible
limitations of the method.  The two most important are time
variability and gravitational microlensing.  Time variability affects
the results through the time delay differences $\Delta t_i$ between
the images.  If the data to be modeled are collected at a single epoch,
then only variations in the shape of the spectrum affect the
extinction results, while if we combine data from multiple epochs both
changes in the total flux and the spectral shape affect the extinction
results.  When time variability is observed in quasars and AGN, it is
seen at all wavelengths simultaneously (e.g. Ulrich, Maraschi \& Megan
1997, Neugebauer et al.\ 1989), so the changes in the overall flux are
generally larger than the changes in the spectral shape.  While the
mean magnification of the lens is constant and achromatic,
gravitational microlensing introduces both temporal variations and
chromaticity (see Schneider, Ehlers \& Falco 1992).  The radio
emitting, optical broad/narrow line, and accretion disk regions have
different physical sizes and thus can have different average
magnifications (see Schneider, Ehlers \& Falco 1992).  Where
variability is observed in lenses it is seen at all wavelengths
simultaneously (e.g. Kundi\'c et al. 1997, Corrigan et al. (1991),
Ostensen et al. (1996)) and in both the continuum and the strong
emission lines (Saust 1994).  Time variability in the mean
magnification is more common and has a larger amplitude than chromatic
variations.

The systematic errors will broaden the extinction distributions beyond
the effects of purely statistical errors, distort extinction curve
estimates, and bias redshift estimates.  The $\chi^2$ values of the
fit provide a simple means of estimating the level of systematic
errors.  If our {\it ansatz} that the spectral differences between the
images are due only to differential magnification and extinction is
correct, then we should find solutions with $\chi^2 \simeq N_{dof} \pm
(2N_{dof})^{1/2}$ where $N_{dof}$ is the number of degrees of freedom.
If we find a poor statistical fit, then the true magnitude errors
produced by the unmodeled systematic uncertainties (time variability
and microlensing) should be of order $(\chi^2/N_{dof})^{1/2}$ larger
than the statistical errors in the magnitudes.  Once the parameters of
the fit were determined, we found the value for the systematic error
parameter $\sigma_{sys}$ which would produce $\chi^2=N_{dof}$ as an
estimate of the level of systematic error required to invalidate the
procedure.

The determination of the total extinction is a more dangerous
procedure.  We will assume that all sources have an intrinsic spectrum
which can be modeled by the mean spectrum of bright optically-selected
quasars.  We produced a composite UV-IR quasar spectrum including the
dominant emission lines by combining the UV-IR continuum model of
Elvis et al. (1994) with the mean spectrum including broad emission
lines of Francis et al. (1991).  We computed the intrinsic colors of
the quasar by convolving the redshifted mean spectrum with the
appropriate filter functions.  By including only lenses dominated by
point sources, we avoid the systematic error in Malhotra et al. (1997) of
interpreting the red colors of extended emission from the source's
host galaxy as reddened emission by a quasar (see Kochanek et
al. 1998b, King et al. 1998).  It is known, however, that the broad
range of colors observed for unlensed, radio-selected quasars is not
exclusively due to contamination by emission from the host galaxy (see
Masci, Webster \& Francis 1998).  The red colors of many
radio-selected sources must be due either to extinction of the AGN by
the host galaxy (e.g.  Wills et al. 1992) or intrinsic differences in
the AGN spectrum (e.g. Impey \& Neugebauer 1988).  Thus, for the
radio-selected lenses we will frequently over-estimate the total
extinction in the lens galaxy because our assumption for the intrinsic
spectrum is either invalid or represents a determination of the amount
of dust in the host galaxy.
   
We first estimated an upper bound to the extinction from the
requirement that the source have an intrinsic magnitude fainter than
$M_B = -29$ mag ($H_0=50$ km~s$^{-1}$~Mpc$^{-1}$).  Brighter
quasars are exceedingly rare at all redshifts (e.g. Hartwick \& Schade
1990).  While we will
not include a correction, an offset would lead to a increase in our
total extinction estimates by the same amount. We estimated the limit
using the bluest image and bluest filter to the red of Ly$\alpha$.  We
used a constant lens magnification correction of a factor of 10 --
using the true magnifications will have only a modest effect on the
limits.  Next we estimated the total extinction from the color of the
bluest image on the longest available wavelength base-line to the red
of Ly$\alpha$.  If we call the red filter $r$ and the blue filter $b$
at observed wavelengths $\lambda_r$ and $\lambda_b$, then

\begin{equation}
     r - b =  (r-b)_0 (z_s)  + E_{Gal}  (R(\lambda_b)-R(\lambda_r)) + 
           E_{lens} (R({\lambda_b \over 1+z_l})-R({\lambda_r \over 1+z_l})) + 
           E_{src}  (R({\lambda_b \over 1+z_s})-R({\lambda_r \over 1+z_s}))
\end{equation}
where $E_{Gal}$, $E_{lens}$ and $E_{src}$ are the $E(B-V)$
contributions from the Galaxy, the lens and the source respectively.
We assume a fixed $R_V=3.1$ Galactic extinction law for all three
components, and corrected the individual lenses for Galactic
extinction using the model of Schlegel, Finkbeiner \& Davis
(1998).\footnote{ The Elvis et al. (1994) model used the Burstein \&
Heiles (1978) Galactic extinction model which may have an offset of
$E(B-V)=0.02$ mag (Schlegel, Finkbeiner \& Davis 1998). }  
The intrinsic color $(r-b)_0 (z_s)$ was derived by redshifting
the mean quasar spectrum and convolving it with the appropriate filter
passbands.  The uncertainties in the colors and extinctions are
entirely systematic.  The spectral index of the optical continuum of
quasars ($F_\nu \propto \nu^\alpha$) is $\alpha \simeq -0.4 \pm 0.6$
(Francis et al. 1991), which would lead to an uncertainty in the
extinction of $\sigma_E \simeq \sigma_\alpha/3 \simeq 0.2$ mag for
typical wavelength baselines and lens redshifts if it also
characterized the uncertainties in the optical through infrared
continuum.  We determine the extinction errors scaled to an
uncertainty in the intrinsic color of $\sigma_{r-b} = 0.1$ mag, and
then estimate the accuracy of the error estimates from the
distribution of extinctions.  Since we have no means of separately
determining $E_{lens}$ and $E_{src}$ given the uncertainties in the
intrinsic spectrum, we have estimated $E_{lens}$ with $E_{src} \equiv
0$.  For red radio quasars created by dust in their host galaxy we
will mistakenly assign $E_{src}$ to $E_{lens}$, and for intrinsically
red radio quasars we will produce nonsense.

\section{Differential Extinction Estimates for $R_V=3.1$ }

\begin{figure}
\centerline{\psfig{figure=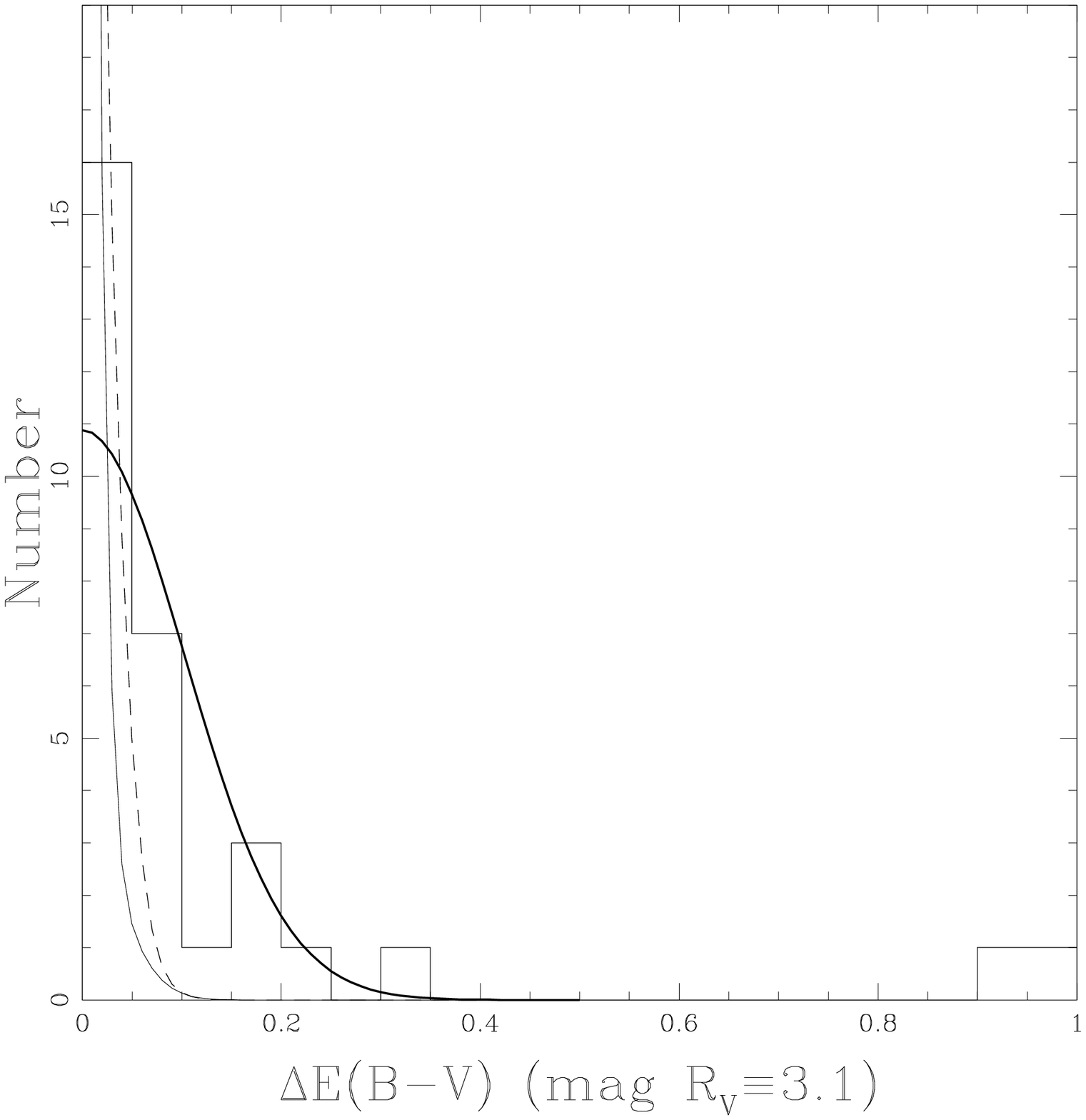,height=6.5in}}
\caption{Histogram of the differential extinction for $R_V\equiv3.1$.
Only values with standard errors smaller than $\sigma(E(B-V)) < 0.1$
mag are included.  The highest extinction lens (PKS~1830--211) is
shown in the rightmost bin, but actually has a higher extinction (see
Table 2).  The light solid curve shows the distribution expected due
to random photometric errors based on the standard errors of the
individual $E(B-V)$.  The light dashed curve shows the expected
distribution due to random photometric errors after rescaling the
errors by $(\chi^2/N_{dof})^{1/2}$ when $\chi^2 > N_{dof}$ to
compensate for underestimated photometric uncertainties.  The light
solid and dashed curves exclude the high differential extinction
lenses B~0218+357 and PKS~1830--211 because their large absolute (but
small fractional) errors are not relevant for the reliability of the
small differential extinction measurements.  The heavy solid curve is
the the best fit Gaussian model for the differential extinctions
excluding B~0218+357 and PKS~1830--211.  }
\end{figure}

\begin{figure}
\centerline{\psfig{figure=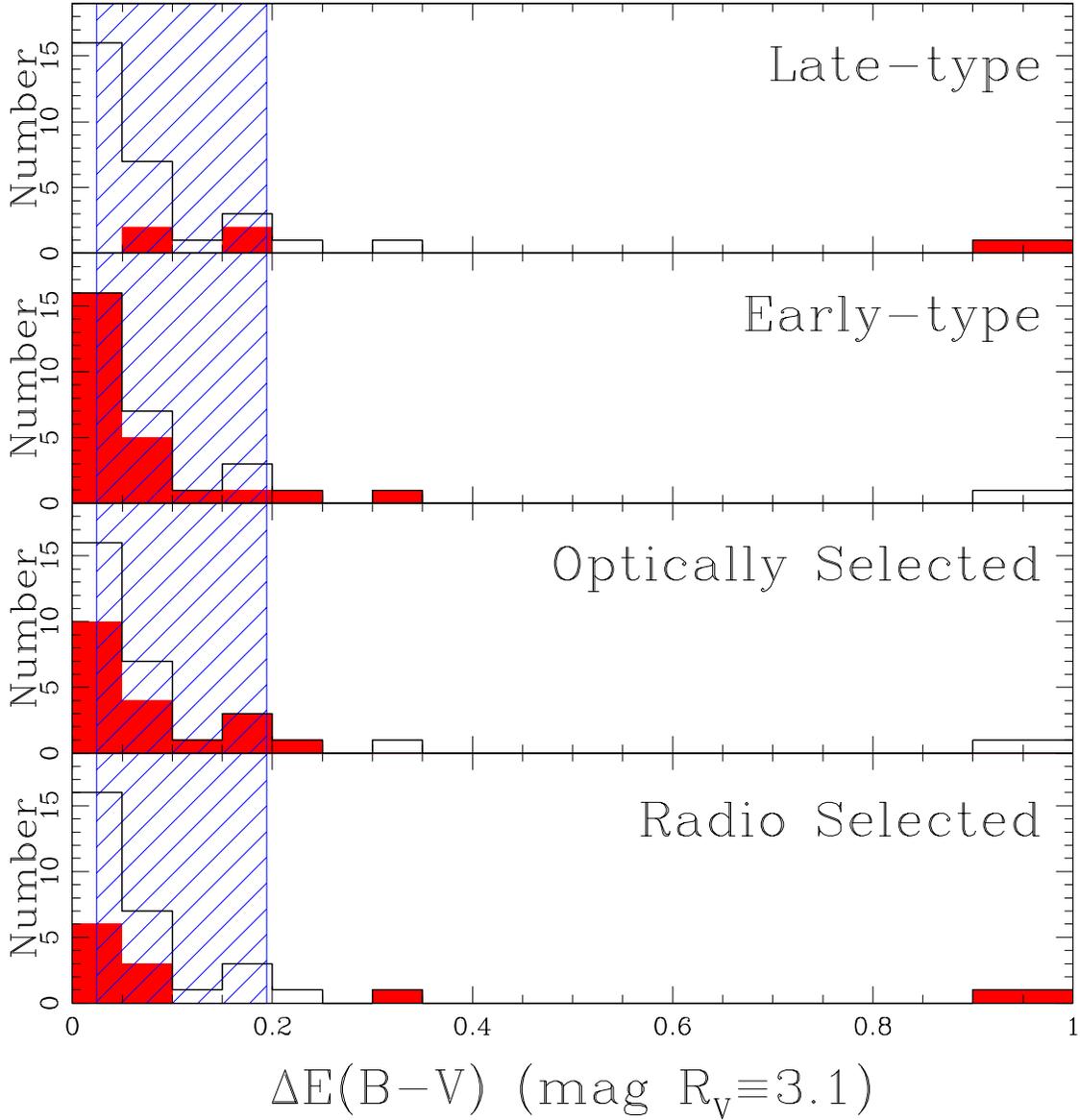,height=6.5in}}
\caption{Histograms of differential extinction estimates by subsample.
The shaded histograms show (from bottom to top) the distributions for
radio-selected, optically-selected, early-type and late-type lenses.
The unshaded histogram superposed on each of the panels shows the
overall distribution from Figure 1. Galaxies with unknown morphologies
have been added to the early-type sample. The hatched region shows the
estimated range for the mean total extinction in lens galaxies
estimated from a comparison of optically-selected and radio-selected
lens statistics by Falco et al. (1998).  }
\end{figure}

\begin{figure}
\centerline{\psfig{figure=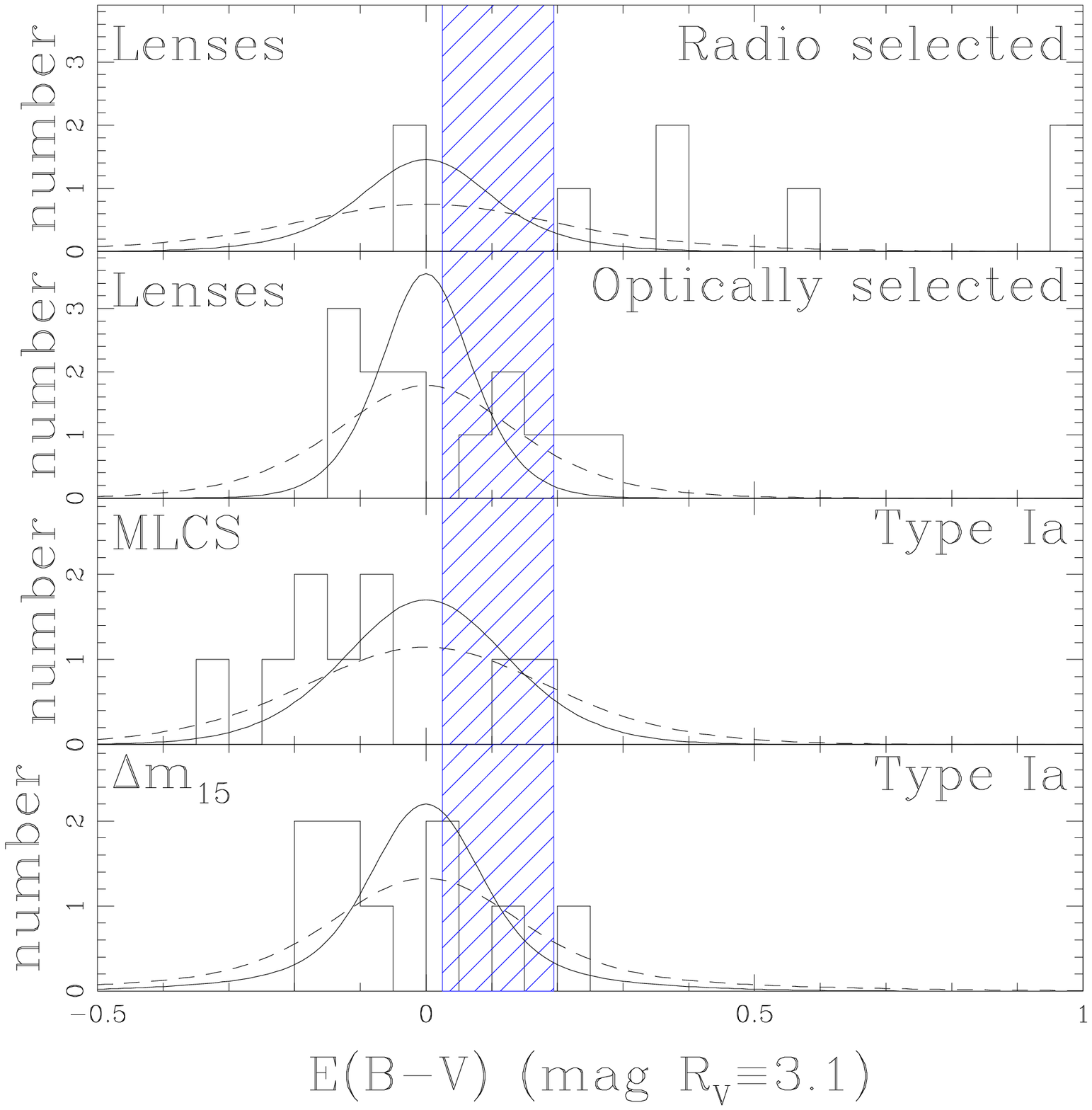,height=6.5in}}
\caption{Histograms of total extinction.  The top histograms show the
total extinction distributions for the bluest images of the
radio-selected and optically-selected subsamples.  The bottom two
histograms show the distribution of Type Ia supernova extinctions from
Riess et al. (1998) using either the MLCS or $\Delta m_{15}$ analysis
methods.  The solid curves show the extinction distributions expected
from random errors given the nominal extinction uncertainties. The
dashed curves show the distributions expected after broadening the
nominal uncertainties by a factor of 2 for the lenses, 1.5 for the
MLCS method, and 1.7 for the $\Delta m_{15}$ method. }
\end{figure}

We start by estimating the differential extinctions in the lenses
because these can be determined independently and more accurately than
the total extinctions.  Differential extinction estimates have been
made for many individual lens systems (e.g. McLeod et al. (1998) most
recently for MG~0414+0534, Hewett et al. (1994) for LBQS~1009--0252,
Chae \& Turnshek (1998) most recently for H~1413+117, Nadeau et
al. (1991) for Q~2237+0305), but here we present the first full,
systematic survey.  We selected 23 lens systems with multiply imaged
point-sources.  Of the 23 systems, 8 were radio-selected (12
differential extinctions) and 15 were optically-selected (25
differential extinctions).  We have fewer radio-selected than
optically-selected systems because so many of the radio lenses are
dominated by extended sources for which our analysis procedure does
not apply.  The exclusion of systems because their sources are
extended is unbiased with respect to the extinction in the lens
galaxy.  Where the lens redshifts are unknown, we have used
preliminary redshift estimates from the fundamental plane of lens
galaxies (Kochanek et al. 1998a).  Modest errors in the redshift
estimates ($\Delta z \ltorder 0.2$) will not significantly alter the
distribution of differential extinctions.  The data we used in the
analysis are presented in Table 1.

We first fit each of the systems using eqn. (3) and assuming no
differential extinction.  Table 2 presents the $\chi^2_1$ statistic
for the fits to each system. We find that 7 of the 23 systems are
consistent with no differential extinction or any other systematic
difference between the spectral shapes of the images.  Next we fit the
systems assuming a $R_V=3.1$ Cardelli et al. (1989) parameterized
extinction curve.  All the initially poorly fit systems show dramatic
improvements in their goodness of fits.  Note, however, that only 10
of the 23 systems have $\chi^2_2$ statistics consistent with a good
fit ($\chi^2_2 \simeq N_{dof}$).  The poor fits can be explained by
underestimated photometric errors (generally by a factor of $\sim 2$),
systematic errors in the extinction calculation (using the wrong
extinction curve or lens redshift), or systematic errors due to
unrelated physics (spectral variability and microlensing).  We attempt
to compensate our quantitative estimates for systematic errors by
rescaling the uncertainties in the extinction or intrinsic flux ratios
by the factor $(\chi^2_2/N_{dof})^{1/2}$ whenever $\chi^2_2 >
N_{dof}$.  The procedure is equivalent to broadening the photometric
uncertainties so that the best fit model has $\chi^2_2=N_{dof}$, or to
redoing the fits with a non-zero value for $\sigma_{sys}$.

The resulting estimates for the differential extinctions and the
intrinsic flux ratios of the images are presented in Tables 3 and 4
using the rescaled uncertainties.  Figure 1 presents the distribution
of differential extinctions $\Delta E(B-V)$ for the overall sample
after dropping the two lenses (HST~12531--2914 and HST~14176+5226)
where the standard error in the estimate exceeds $\sigma(\Delta
E(B-V))>0.1$ mag.  Photometric errors would produce a distribution of
non-zero extinctions even in the absence of dust.  We can estimate
this ``null'' distribution by the normalized sum of one-sided
Gaussians for each measurement (excluding B~0218+357 and PKS~1830--211
whose large errors absolute but small fractional errors are not relevant
to the null distribution), leading to the very narrow distribution
contained within the first bin of Figure 1.  If we use the rescaled
errors instead of the original errors, the null distribution is
broader but still significantly narrower than the observed
distribution.  Thus, the differential extinction distribution is only
marginally broadened by the effects of photometric uncertainties.  If
we fit the distribution of differential extinctions excluding
B~0218+357 and PKS~1830--211 with a Gaussian broadened by either the
standard or rescaled statistical error estimates we find a best fit
width for the distribution of $\sigma_{\Delta E}=0.1$ mag, which is
roughly consistent with the observed median of $0.05$ mag (see Figure
1).

Figure 2 presents the extinction distributions for two different ways
of dividing the data into subsamples.  The first division is into
radio-selected and optically-selected lenses, where the first is
unbiased and the second is biased with respect to the amount of
extinction.  The second division is into early-type and late-type
lenses.  Four of the lenses (the optically-selected lens Q~2237+0305,
and the radio-selected lenses B~0218+357, B~1600+434, PKS~1830--211)
are late-type galaxies based on morphology, color, or molecular gas
content.  Most lenses are early-type galaxies which lie on the fundamental 
plane (Kochanek et al. 1998a) with the the expected colors, sizes and 
morphologies (Keeton et al. 1998).  Since most lenses are expected to 
be early-types, we have put the currently unclassified lenses (SBS~0909+532, 
LBQS~1009--0252, Q~1017--207, Q~1208+1011, and H~1413+117) into the early-type
subsample.  It is not surprising that the two
highest-extinction systems, B~0218+357 and PKS~1830--211, are both
radio-selected systems and have late-type lens galaxies.  Both are
nearly face-on, cover the quasar images, have exponential rather than
early-type photometric profiles (Leh\'ar et al. 1999) and contain
large amounts of molecular gas (see \S4).  The other two late-type
galaxies lie amidst the main lens population.  For Q~2237+0305 we see
the images through the bulge rather than the disk of the galaxy, and
for B~1600+434 the images lie on either side of the edge-on lens galaxy.
The median extinction of the optically-selected subsample of $\Delta
E(B-V)=0.04$ mag is slightly smaller than the median extinction of the
radio-selected sample of $\Delta E(B-V)=0.06$ mag, as might be
expected given the bias against dusty systems in the optical samples.

In the four-image systems the images usually all lie at a common
distance from the lens center, while in the two-image systems the
images usually lie at significantly different distances from the image
center.  We find no pattern in the extinction as a function of impact
parameter relative to the lens center.  In half of the lenses the
inner image is more reddened and for the other half of the lenses the
outer image is more reddened.  For example, the inner image of
B~0218+357 passes $0\farcs1$ from the lens center and the outer image
passes $0\farcs2$ from the lens center, but the outer image shows
$\Delta E(B-V)=0.9 $ mag more extinction than the inner image.  Thus,
the dust distribution in the lens galaxies must be patchy, just as in
nearby galaxies.

\section{Estimates of the Total Extinction for $R_V=3.1$ and 
the Dust-to-Gas Ratio}

We discuss the estimates of the total extinction separately because
the need to assume an intrinsic color for the quasars makes the
results less reliable than the estimates for the differential
extinction.  The estimates of total extinction are also presented in
Table 3, along with the upper bound on the rest-frame $A_V$ above
which the source luminosity exceeds that of the brightest known
quasars.  In most cases the extinction estimated from the image colors
is significantly below the physical limit set by the maximum quasar
luminosities.  The median total
extinction of the bluest images is $E(B-V)=0.08$ mag.  We can also
estimate the typical total extinction of the lens systems by comparing
the statistics of radio and optical lens surveys.  If we normalize the
statistical models using the radio surveys, then we can use the
optical surveys to estimate the mean extinction needed to reconcile
the two samples.  Falco et al. (1998) performed such a comparison and
found $\langle A_B \rangle\simeq 0.58 \pm 0.45$ mag in the observers
rest frame. For a typical lens redshift of $z_l \simeq 0.5$ the B band
extinction coefficient at the lens is $R_{QSO} \simeq 6$ and thus the
mean extinction is $\langle E(B-V)\rangle \simeq \langle
A_B\rangle/R_{QSO} \simeq 0.10 \pm 0.08$ mag.  The estimated
median total extinction for the individual lenses agrees with the
extinction estimated from the statistical comparison.

Figure 3 shows the extinction distributions for the blue images
divided into radio-selected and optically-selected subsamples.  The
superposed curves show the distribution predicted from random errors
in the intrinsic source color using either the standard error
($\sigma_{r-b}=0.1$ mag) or twice the standard error.  The absence of
a broader tail of negative extinctions means that our standard error
estimate is approximately correct, and that we are at most
underestimating the uncertainties by a factor of two.  Even so, much
of the total extinction distribution, particularly for the
optically-selected subsample, is dominated by the uncertainties in the
intrinsic colors.  We also see no correlation between the differential
and total extinctions for the low total extinction lenses ($E(B-V)
\ltorder 0.2$ mag), which is probably a sign that the distribution is
dominated by noise.

It is clear that the distribution of the radio-selected lenses differs
from that of the optically-selected lenses.  We know that
radio-selected quasars have a broader range of colors than
optically-selected quasars (e.g. Webster et al. 1995), and that
contamination from the host galaxy can be only a partial explanation
(e.g. Masci et al. 1998).  In the two high differential extinction
lenses, B~0218+357 and PKS~1830--211, it is likely that the total
extinctions of $E(B-V)=0.6$ mag can be attributed to dust in the
lens. On the other hand, MG~0414+0534 has the highest total extinction
estimate for the blue image ($E(B-V)=1.4$ mag or 5 times the largest
differential extinction in the system) but the extinction is probably
due to dust in the source.  A model using only dust in the lens must
explain why there are $ < 20\%$ variations in the extinction across a
region spanning $13 h_{65}^{-1}$ kpc, and that the arc image of the
quasar host galaxy is significantly bluer than the quasar images
(Falco et al. 1998). The intrinsic spectra of galaxies are redder than
those of optically-selected quasars, so the only way to find a blue
arc and a red quasar is to put the dust in the source.  Dust in the
lens would produce a red quasar and a still redder arc.  Moreover, the
inferred intrinsic luminosity of the arc is $\sim 0.5 L_*$ without any
extinction correction, so adding $A_B=6$ mag of dust in the lens is
implausible.  We found a similar situation in the lens MG~1131+0456
(Kochanek et al.  1998b), where the lens galaxy is demonstrably
transparent but dust in the host galaxy obscures the AGN cores.  Thus,
the broad range of extinctions estimated for the radio-lenses is
problematic.

Absorption by HI and molecular gas in the lens galaxy is seen in two
systems, B~0218+357 and PKS~1830--211.  For these systems we can use
our direct estimates of the extinction to determine the dust-to-gas
ratios of the lens galaxies.  In both systems the molecular and atomic
absorption is seen in front of only one image.  In B~0218+357 the
molecular gas at $z_l=0.69$ is seen in front of the A image (Menten \&
Reid 1996). In PKS~1830--211 the molecular gas at $z_l=0.89$ is seen
in front of the B (Southwest) image (Frye, Welch \& Broadhurst 1997),
and there may be HI absorption at $z=0.19$ in front of the A
(Northeast) image (Lovell et al. 1996).  The molecular absorption is
inferred to be in optically thick CO clouds which incompletely cover
the continuum radio source.  For B~0218+357 the H$_2$ column density
estimates range from $N(H_2) \simeq 2 \times 10^{22}$ cm$^{-2}$
(Wiklind \& Combes 1995, Gerin et al. 1997) to $5 \times 10^{23}$
cm$^{-2}$ (Wiklind \& Combes 1995, Combes \& Wiklind 1997), combined
with an HI column density of $N(HI) \simeq 10^{21}$ cm$^{-2}$
(Carilli, Rupen \& Yanny 1993). For PKS~1830--211 the H$_2$ column
density in front of the B image is estimated to be $N(H_2) \simeq 2.5
\times 10^{22}$ cm$^{-2}$ (Wiklind \& Combes 1996, 1998, Gerin et al.
1997), and the HI column density in front of the A image is estimated
to be $N(HI) \simeq 10^{20}$ cm$^{-2}$.  X-ray spectra of
PKS~1830--211 show a deficit of soft X-rays compared to normal quasar
X-ray spectral indices, which could be explained by a gas density in
the lens of $N(H) \simeq 3.5 \times 10^{22}$ covering both images 
(Mathur \& Nair 1997).

For both lenses, standard conversions of gas surface density to
extinction (e.g. $A_V \simeq 6.3 \times 10^{-22} \left[ N(HI) + 2
N(H_2)\right]$, Savage et al. 1977)) predict total extinctions of $A_V
\gtorder 20 $ mag.  Such an estimate is patently wrong, because for
$A_V \gtorder 20$ the quasars would be unphysically luminous.
The luminosity limit of $M_B > -29$
mag sets an upper bound to the total extinction of the blue image of
$A_V < 3.5$ for B~0218+357 and $A_V < 4.1$ for PKS~1830--211.  The
extinction estimates for B~0218+357 are $A_{VA}=4.7$ mag and
$A_{VB}=1.9$ mag, corresponding to molecular column densities of
$N(H_2) \simeq 4 \times 10^{21}$ cm$^{-2}$ and $ 2 \times 10^{21}$
cm$^{-2}$ respectively.  The extinction estimates for PKS~1830--211
are $A_{VA} = 1.8$ mag and $A_{VB} = 11.1$ mag respectively,
corresponding to molecular column densities of $N(H_2) \simeq 1 \times
10^{21}$ cm$^{-2}$ and $ 9 \times 10^{21}$ cm $^{-2}$ respectively.
In both cases, either the dust-to-gas ratio is a factor of 2--5 times
lower than standard estimates, or the optical source coincidentally
lies behinds one of the gaps in the molecular gas coverage.  The low
extinction of the PKS~1830--211~A image probably means that there is
little X-ray absorption by the lens at the A image and that Galactic
absorption must be largely responsible for suppressing the soft
X-rays.  The high $R_V$ extinction curves that best fit the two lenses
(see \S5) alter the differential component of $A_V$ from $2.8$ to
$3.5$ mag for B~0218+357 and from $9.3$ to $7.2$ mag for PKS~1830-211,
which is insufficient to explain the apparent change in the
dust-to-gas ratio.

\section{Extinction Laws and Lens Redshifts }

Determining extinction laws (Nadeau et al. 1991) and lens redshifts
(Jean \& Surdej 1998) is more difficult than determining extinctions
because they are more sensitive to the systematic errors introduced by
variability and microlensing.  
The current data are not ideal because
the wavelength coverage is usually limited to the V band and redder
filters, and because we are forced to combine data from epochs
spanning a period of several years.  The results of this section
should be regarded more as a feasibility study rather than a final,
complete survey.  In particular, the reliability would be considerably
increased if the data were obtained at a common epoch.  Note, however,
that we are again using only the spectral differences between the images
and avoid the uncertainties about the intrinsic spectrum.

We first found the best fit value of $R_V$ for the Cardelli et
al. (1989) parameterized extinction curves using fixed lens redshifts
and a prior probability of $R_V=3.1\pm1.0$ on the extinction law.  We
adopt the prior to stabilize the results when the data are inadequate
to determine $R_V$.  Table 3 presents the goodness of fit,
$\chi^2_3/N_{dof}$, and the estimated value of $R_V$.  We can divide
the systems into three loose categories.  The first group consists of
the seven lenses with small differential extinctions ($\Delta E(B-V) <
0.04$ mag, Q~0142--100, BRI~0952--0115, Q~0957+561, Q~1017--207,
B~1030+074, PG~1115+080, HE~2149--2745) and the two lenses with large
fractional uncertainties in the differential extinction
(HST~12531--2914, HST~14176+5226).  For these systems the extinction
curve is either poorly determined or overly subject to small
systematic errors.  For example, Q~0957+561 has little extinction
($\Delta E(B-V) = 0.02\pm0.02$ mag for $R_V=3.1$), but we derive a
value of $R_V=7.7\pm0.9$. The large change in the $\Delta \chi^2$
between fixed and variable $R_V$ is almost certainly due to using
photometric uncertainties which are underestimates of the true
uncertainties (i.e. $\sigma_{sys} > 0$).  The second group consists of
SBS~0909+532 and B~1600+434 for which we derive $R_V < 1$.  We believe
these are unphysical and the result of problems in the data.  The
SBS~0909+532 photometry is mainly from ground-based observations that
may be significantly contaminated by the lens galaxy, while the
optical and radio flux ratios for B~1600+434 are poorly determined and
variable.

The third group consists of the 12 systems with plausible $R_V$
estimates. Seven are consistent with $R_V=3.1$ given the uncertainties
(LBQS~1009--0252, HE~1104--1805, Q~1208+1011, H~1413+117, B~1422+231,
SBS~1520+530 and MG~2016+112) and five are not (B~0218+357,
MG~0414+0534, FBQ~0951--0115, PKS~1830--211 and Q~2237+0305).
However, for none of the systems consistent with $R_V=3.1$ can the
existing data significantly reduce the uncertainties beyond the level
specified by the prior.  Added data can improve little the results for
H~1413+117 and B~1422+231 where the differential extinctions are
fairly low and there is good wavelength coverage in the existing data.
The other systems have larger extinctions and poorer wavelength
coverage so the prospects for significantly reducing the extinction
curve uncertainties are good.  Of the five with atypical values of
$R_V$, three are variable (B~0218+357--Biggs et al. 1998 (radio),
Wagner 1998 (IR); PKS~1830--211--Lovell et al. 1998 (radio);
Q~2237+0305--Corrigan et al. 1991, Ostensen et al. 1996 (optical)),
and one is not variable (MG~0414+0534--Moore \& Hewitt 1997 (radio),
Schechter 1998 (optical)).  There are no variability data on
FBQ~0951--0115 or IR--UV variability data for PKS~1830--211.

For B~0218+357, MG~0414+0534, MG~2016+112 and Q~2237+0305 we can
derive non-parametric extinction curve estimates with sufficiently
small formal uncertainties to compare to the parametric
determinations.  Figure 4 shows the standard ($R_V=3.1$) and best fit
parametric models along with the non-parametric model. The
non-parametric extinction curves match the best fit parametric curves
given the formal uncertainties for three of the systems, while for
MG~2016+112 there is a discrepant point associated with the Gunn r
band photometry.  Note that distinguishing the best fit extinction
curve from the typical Galactic $R_V=3.1$ curve depends almost
entirely on the data to the blue of the rest frame V band.  With only
data to the red of the rest frame V band, the self-similarity of the
extinction curve makes it almost impossible to distinguish the models.
All four lenses are radio sources, so the agreement of the
non-parametric and parametric models rules out the existence of a
significant gray dust opacity over the IR--UV wavelength range.  The
high $R_V$ extinction curves are, however, relatively gray compared to
the standard $R_V=3.1$ curve.

Three of the four systems have significantly non-standard extinction
curves, so we need to demonstrate that the results are due to
extinction rather than systematic error.  One test is to show that the
extinction curve is not varying with time.  For example, the
Q~2237+0305 curve includes time averaged ground-based light curves at
V, R and I from Ostensen et al. (1996) covering epochs from 1990 to
1995 as well as the V and R HST observations from 1995 and 1990
respectively.  We processed the data as separate points in the
extinction curve and found that the $R_\lambda$ values derived from
the ground-based and HST data agree given their mutual uncertainties.
Note, however, that if we include the earlier Corrigan et al. (1991)
ground-based photometric data, we obtain discordant results for the
extinction curve.  Whether this is an artifact of variability or a
consequence of the more sophisticated image modeling approach of
Ostensen et al. (1996) over Corrigan et al. (1991) is unclear.
  
\begin{figure}
\centerline{\psfig{figure=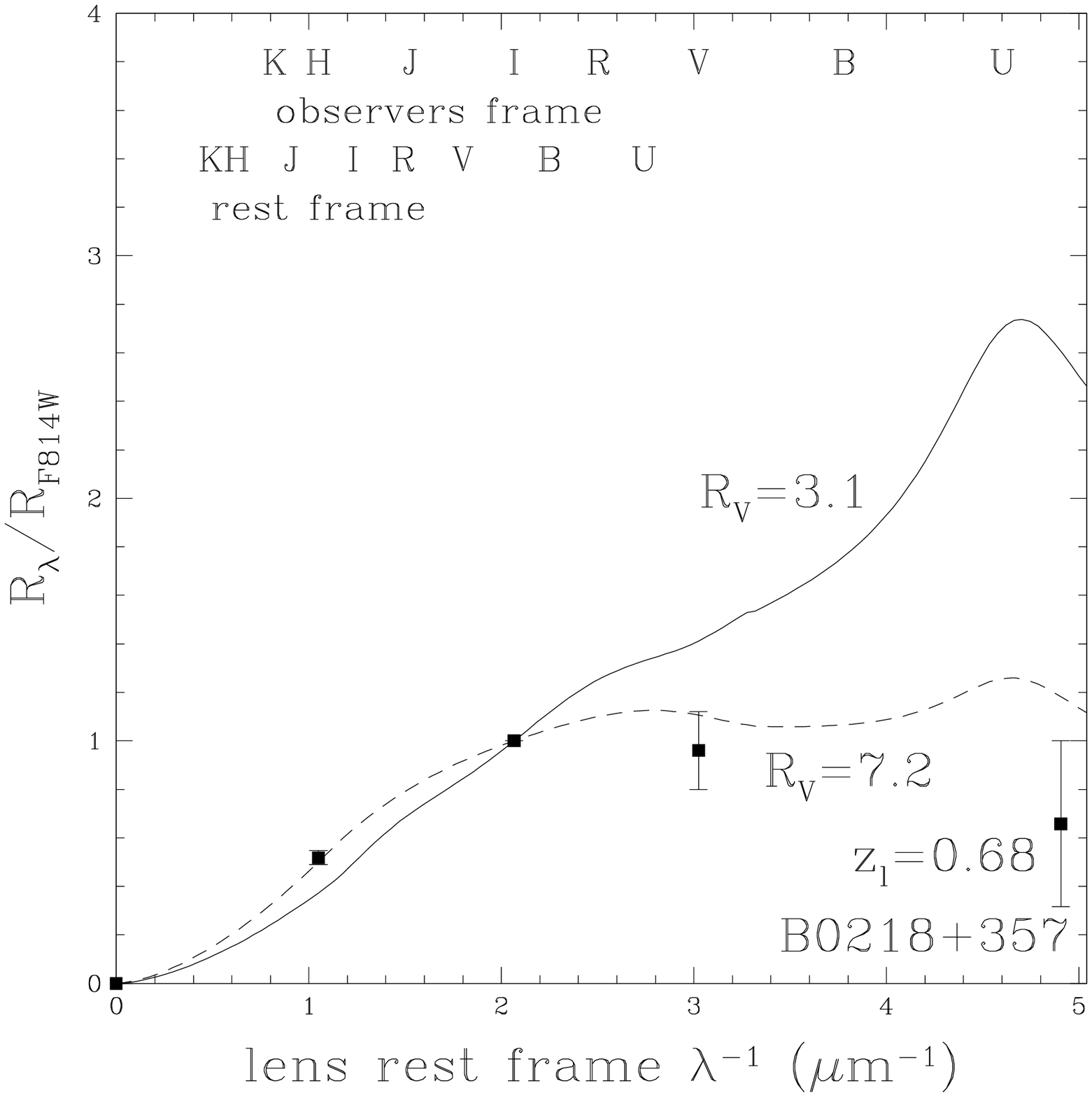,height=3.5in}
            \psfig{figure=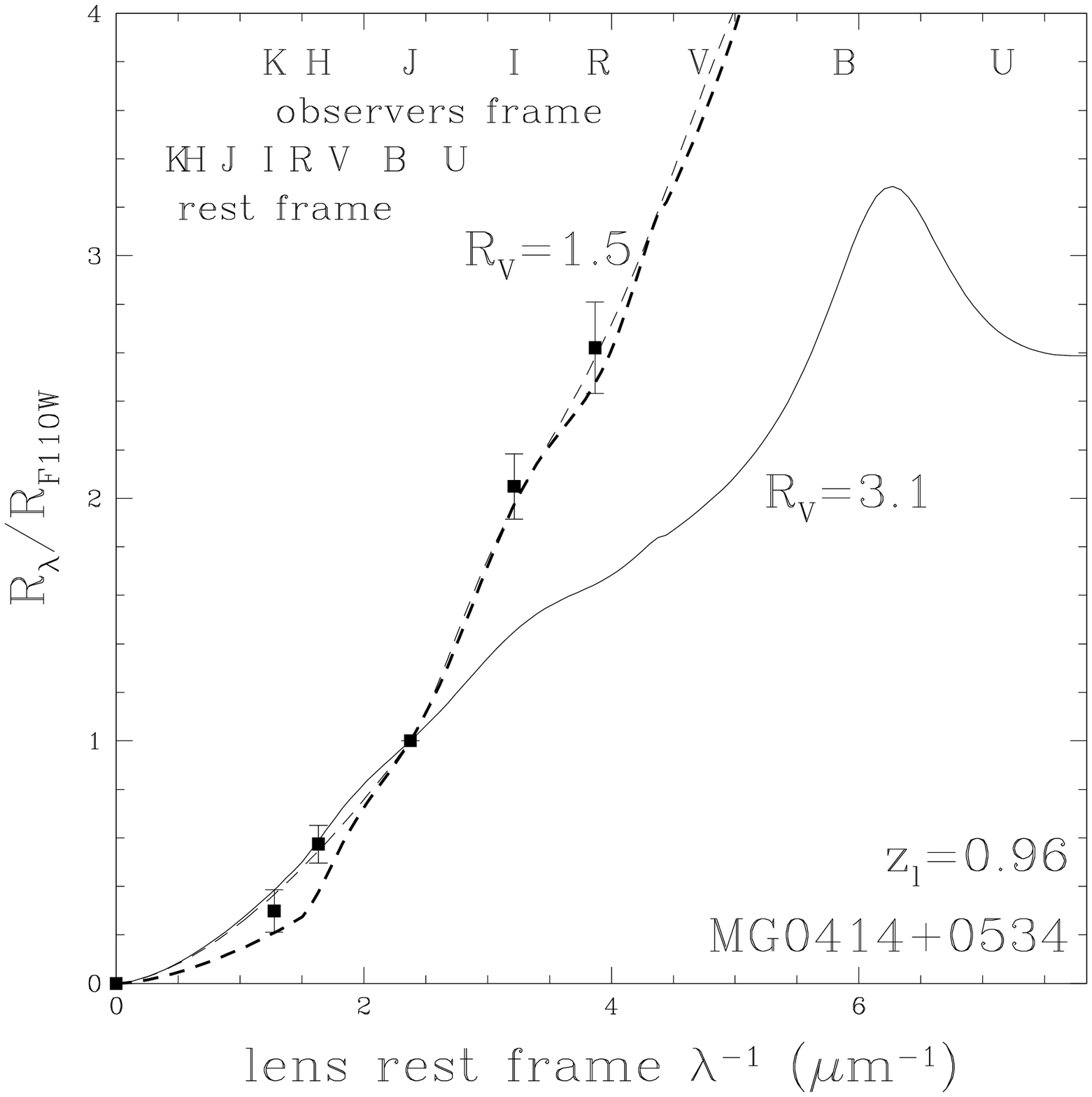,height=3.5in}}
\centerline{\psfig{figure=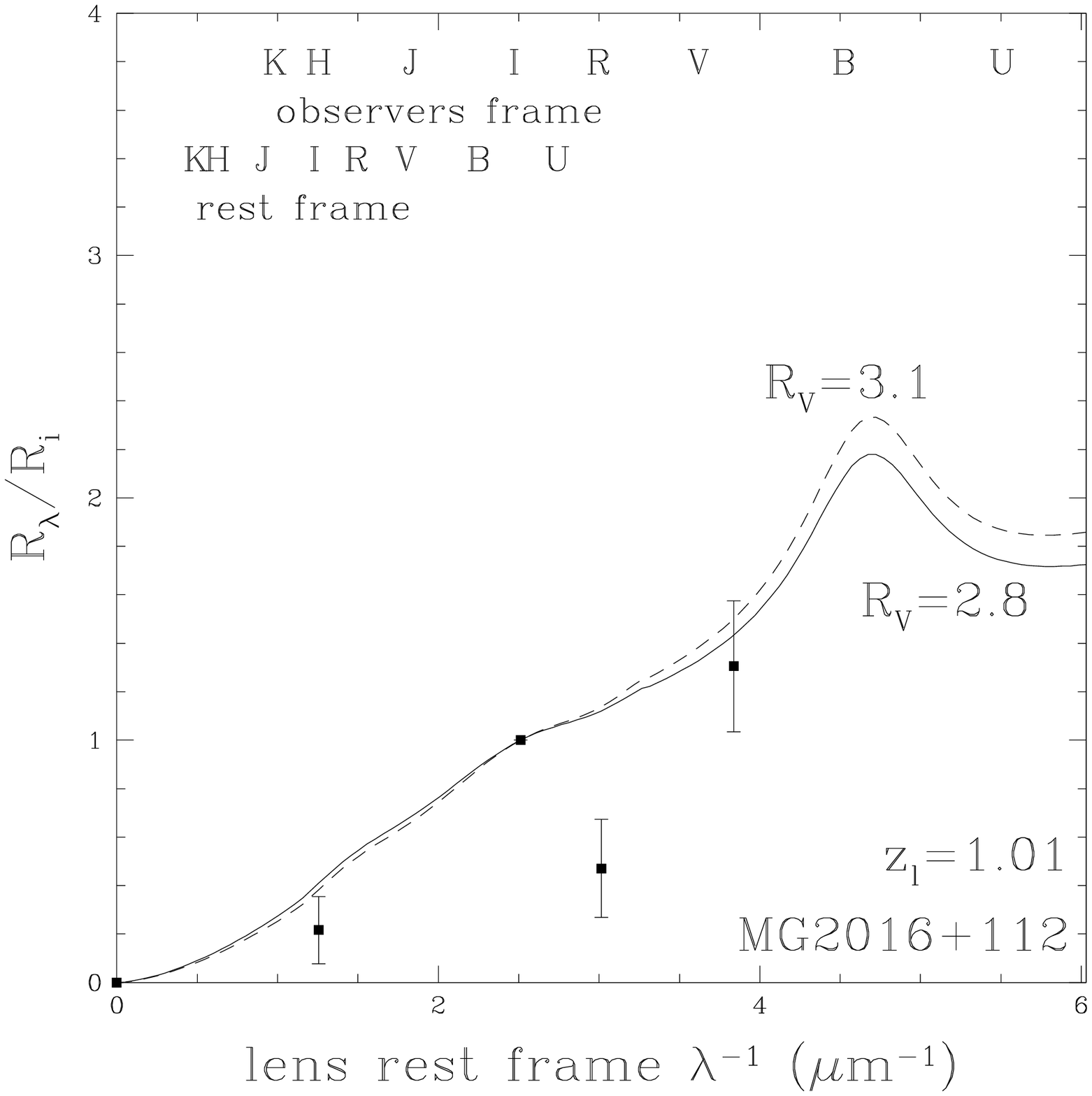,height=3.5in}
            \psfig{figure=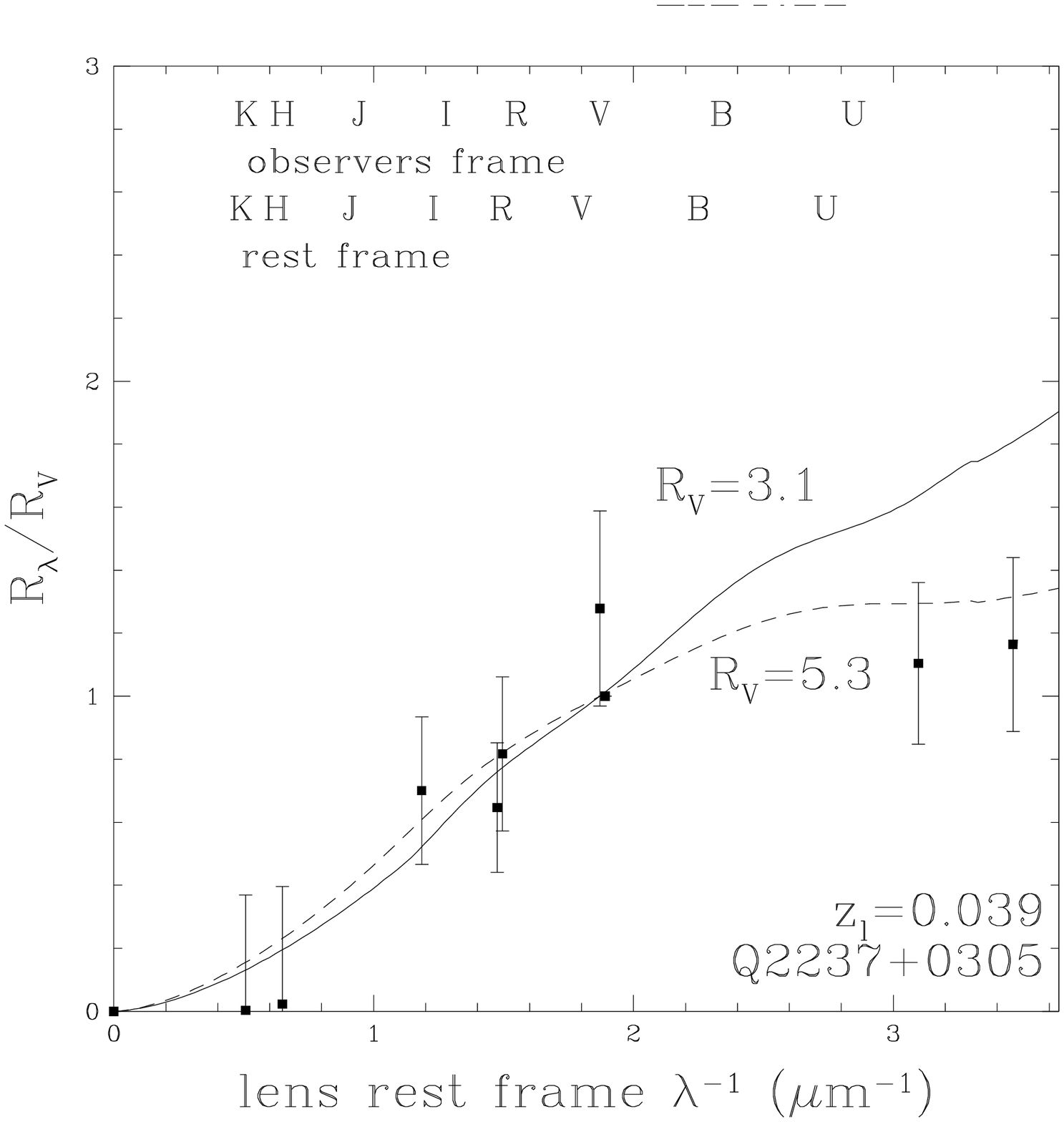,height=3.5in}}
\caption{Extinction curves for B~0218+357 (top left), MG~0414+0534
(top right), MG~2016+112 (bottom left) and Q~2237+0305 (bottom right).
The solid line shows the standard $R_V=3.1$ curve and the dashed line
shows the best fit parametric curve.  For simplicity the curves are
normalized by the $R_\lambda$ value of the filter closest to the lens
rest frame V band.  All four lenses are radio sources and include a
point at $\lambda^{-1}=0$, allowing an absolute determination of both
the extinction and the extinction curve.  For MG~0414+0534 the heavy
dashed line is an $R_V=1.5$ extinction curve at the true lens redshift
of $z_l=0.96$, while the light dashed line is the better fitting
$R_V=2.1$ extinction curve at $z_l=0.23$, thereby illustrating the
dust redshift degeneracy. }
\end{figure}

\begin{figure}
\centerline{\psfig{figure=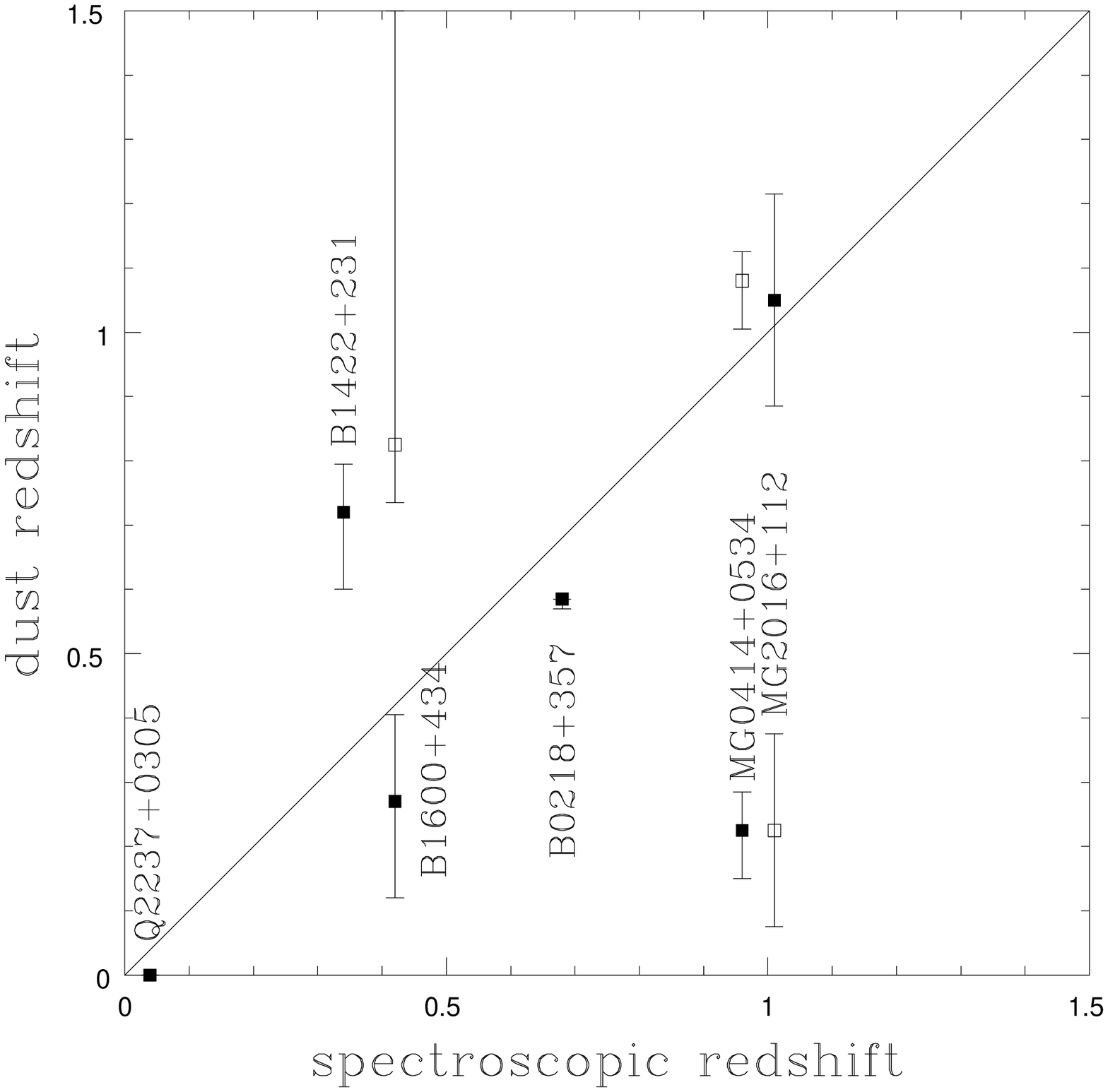,height=6.5in}}
\caption{A comparison of spectroscopic and dust redshifts.  The
redshift uncertainties were determined by the region with
$\Delta\chi^2=1$ from the minimum. The filled points are the primary
(deepest) minima, and the open points are secondary minima in the
$\chi^2$ surface. Figure 4 shows the similarity of the two best fit
extinction curve-redshift solutions for MG~0414+0534.}
\end{figure}

Next, for the systems where the extinction curve was determinable at a
single redshift we attempted to estimate the redshift and the
extinction law simultaneously.  Four problems potentially limit the
determination of dust redshifts.  First, the extinction law in the IR
and NIR is almost self-similar, $R_\lambda \propto
\lambda^{1.7\pm0.1}$, and only the deviations from self-similarity
permit a redshift determination.  Second, lines of sight in the Galaxy
with the same value of $R_V$ do not have exactly identical extinction
curves (e.g. Mathis \& Cardelli 1992).  The value of $R_V$ is the most
important parameter of the extinction curve, but it is not the only
parameter.  Third, the largest deviations from self-similarity occur
in the UV and are associated with the behavior of the 2175\AA\ bump.
However, it is the structure of the UV extinction curve that varies
the most between the Galaxy and the SMC.  Fourth, at some level the
lensing data must be contaminated by microlensing and the resulting
wavelength dependence of the magnification.  These systematic errors
will be far more important for dust redshift determinations, which
depend on the small deviations from a self-similar extinction curve,
than for estimates of $E(B-V)$.  Thus, estimates of the accuracy of
dust redshifts based on Monte Carlo simulations of photometric errors,
as considered by Jean \& Surdej (1998), are likely to underestimate
the true uncertainties.  We will take a purely empirical approach, and
determine the dust redshifts for every system with adequate data and
then compare the known and estimated redshifts.

In Table 5 we present the dust redshift estimates for all lenses with
$E(B-V) > 0.05$ mag and standard errors $\sigma(E(B-V)) < 0.1$ mag for
$R_V=3.1$.  We also included B~1422+231 which has slightly less
extinction because the lens redshift is known and can be compared to
the dust redshift.  Of the 12 lenses, three yielded no good redshift
estimate (SBS~0909+532, HE~1104--1805, and SBS~1520+530) in the sense
that the redshift range covered by the standard errors in the redshift
estimate were too large ($\Delta z_{dust} > 0.5$).  Four of the 
lenses had two minima in the $\chi^2(z_s)$ curve, and we present both
solutions for the redshift.  Figure 5 compares the dust redshifts and
their standard errors for the six lenses which also have spectroscopic
redshifts (B~0218+357, MG~0414+0534, B~1422+231, B~1600+434,
MG~2016+112 and Q~2237+0305).  In four cases the primary (deepest)
minimum is close to the true redshift, and in two cases it is not.
MG~0414+0534, B~1600+434 and MG~2016+112 have both primary and
secondary solutions, one of which lies near the true lens redshift.
Note, however, that the primary minimum for MG~0414+0534 lies near
$z_{dust}=0.20$ instead of the true lens redshift of $z_l=0.96$ with
$\Delta \chi^2=7.1$ (some of the difference is due to the the
prior on $R_V$).  The high and low redshift solutions have
extraordinarily similar extinction curves (see the MG~0414+0534 panel
of Figure 4), and the relative probability of the two solutions is
controlled by the systematic uncertainties in the flux ratios and the
extinction curve.  It is clear that simple Gaussian statistics based
on the photometric errors are not a reliable means of discriminating
between primary and secondary minima.  We find a dust redshift
consistent with the true redshift for five of the six cases in the
sense that $|z_{dust}-z_l|$ is either consistent with the (Gaussian)
uncertainties or exceeds an absolute accuracy of $|z_{dust}-z_l|<
0.1$.  The exception is B~1422+231, which has the least extinction of
the twelve.  If we drop B~1422+231 for having insufficient extinction
and consider only the correct minima, the absolute accuracy of the
dust redshifts is good, with $\langle
z_{dust}-z_l\rangle=0.00\pm0.08$.  The formal redshift uncertainties
are reasonable for MG~0414+0534, B~1600+434 and MG~2016+112, but are
clearly underestimates for Q~2237+0305 and B~0218+357.  In summary,
dust redshift estimates for lenses with sufficient differential
extinction appear to accurately determine the lens redshift, but can
produce multiple, degenerate solutions that cannot be safely
distinguished using simple Gaussian statistics.

\section{Consequences for Cosmology}

Estimates of the cosmological model using both gravitational lenses
and Type Ia supernovae are very sensitive to extinction.  For
optically selected gravitational lens samples, extinction reduces the
amount of magnification bias (see Kochanek 1996), and for supernovae
it increases the inferred distance modulus in the standard fashion.
For both methods, the inferred matter density is underestimated by
approximately $\Delta \Omega \simeq -A_B$ in magnitudes for flat
cosmological models.

Falco et al. (1998) estimated the mean effects of extinction by
comparing the statistics of optically-selected quasar and radio lens
samples.  They found that the two samples could be reconciled if the
mean extinction of a lensed quasar was $\langle A_B \rangle \simeq 0.6
\pm 0.4$ mag or $E(B-V) \simeq 0.10 \pm 0.08$ mag.  As we discussed in
\S3 and \S4 and illustrated in Figures 2 and 3, the directly observed
lens extinction distributions are consistent with the statistical
estimates.  In Figure 6 we illustrate the changes in the probability
of finding a lensed quasar for several simple models of the extinction
distributions.  We simplified the calculations by fixing the
conversion from $E(B-V)$ to $A_B$ in the observer's frame to that for
a lens at $z_l=0.5$ with an $R_V=3.1$ extinction curve.  Otherwise the
calculations follow those of Kochanek (1996). We first considered
analytic models using a two-sided Gaussian distribution of
differential extinctions (i.e. no correlation between impact parameter
and extinction) of width $\sigma_{\Delta E}$ and a one-sided Gaussian
distribution of total extinctions of width $\sigma_E$.  In \S3 we
found that $\sigma_{\Delta E} \simeq 0.1$ mag excluding the two high
differential extinction lenses, and in \S4 we found that the median
total extinction was $0.08$ mag. The total extinction estimate is
inaccurate because of the large uncertainties in the intrinsic source
colors.  We also simply use the measured extinctions as a model of the
distribution, without any corrections for the broadening of the
distribution by measurement errors other than to reset the negative
total extinctions to be zero.  We made the calculation using either
the differential or the combined total and differential extinction
distributions for either the radio or optically selected lenses.
 
Small amounts of differential extinction have little effect on the
statistics (see Figure 6).  A Gaussian differential extinction
distribution with $\sigma_{\Delta E}=0.1$ mag lowers the expected
number of lensed quasars by about 20\%.  We get the same result if we
use the observed differential extinction distribution for the
optically-selected lens sample.  If we use the differential extinction
distribution of the radio-selected lens sample, the expected number of
lenses drops by 25--35\% because of the two radio lenses with high
differential extinctions which were not included in the Gaussian
model.  The total extinction has a much stronger effect on the lens
probabilities -- for $\sigma_E=0.1$ mag the expected number of lenses
drops by 40--60\% if $\sigma_{\Delta E}=0.1$ mag and by 30--50\% if
$\sigma_{\Delta E}=0.05$ mag.  Similarly, the number of lenses drops
by 40--50\% if we use the observed total and differential extinctions
for the optically-selected sample, and by 80\% if we use the
distribution for the radio lenses.  The 80\% drop is inconsistent with
the Falco et al. (1998) comparison of the statistics of the two types
of lens samples, and it provides further evidence that the intrinsic
colors of radio sources are different from those of bright 
optically-selected quasars.

\begin{figure}
\centerline{\psfig{figure=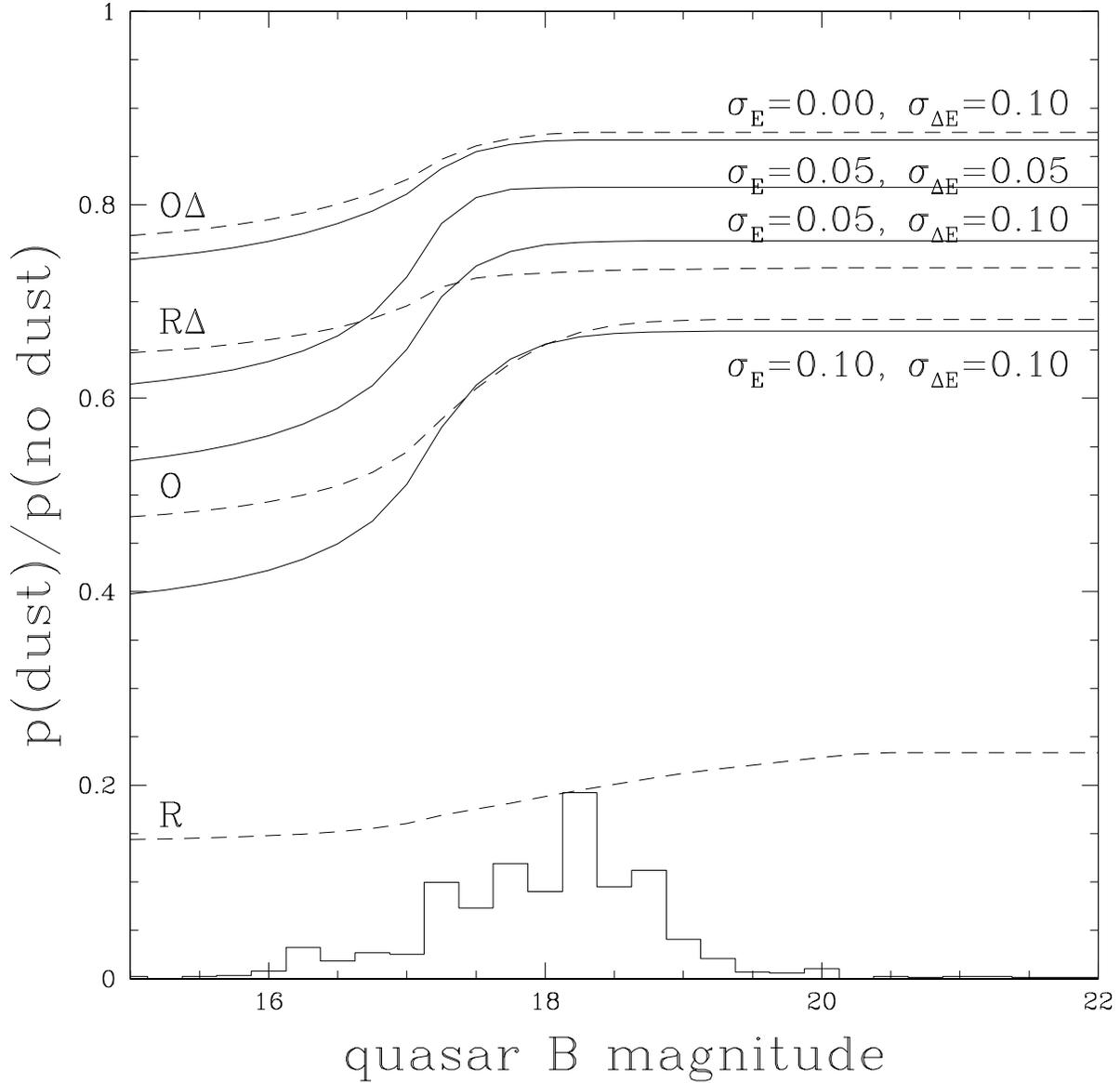,height=6.5in}}
\caption{Effects of extinction on the probability of lensing quasars.
The histogram shows the magnitude distribution of quasars surveyed for
gravitational lenses from Kochanek (1996).  The solid curves show the
ratio of the probability of lensing with dust to that without dust as
a function of quasar magnitude for four analytic models.  In the
analytic models (solid curves) the dust is characterized by the
dispersions of the (one-sided) Gaussian distributions for the
extinction of the bluest image, $\sigma_E$, and the differential
extinction $\sigma_{\Delta E}$.  In the empirical models (dashed
curves) we use the observed distributions for the radio (R) and
optical (O) samples using either only the differential extinction
(R$\Delta$, O$\Delta$) or both the total and differential extinctions
(R, O).  Negative total extinctions were simply set to zero, and no
corrections are made for broadening of the distribution by photometric
errors.  }
\end{figure}

The lens galaxies lie at redshifts comparable to those of the host
galaxies of the Type Ia supernovae used by Riess et al. (1998) and
Perlmutter et al. (1997) to determine the cosmological model.  The
impact parameters of the lensed images are also comparable to those of
the supernovae.  Thus, the extinctions seen in the lens galaxies must
to some extent be comparable to those expected for the supernovae.
The two samples are not, however, identical. The lenses accurately
determine the extinction differences and poorly determine the total
extinction on paths passing completely through the lens galaxy, while
the supernovae are sensitive only to the total foreground extinction.
The lens galaxies are dominated by massive, metal-rich, early-type
galaxies, while the supernovae host galaxies are typically lower mass,
lower metallicity, late-type galaxies.  Later type hosts would imply
more dust, lower metallicity would imply less dust, and total
foreground extinction rather than differential extinction could imply
either more or less dust.

We can make three qualitative statements about what should be observed
for the supernovae given the results for extinction in gravitational
lenses.  First, we clearly find evidence for patchy extinction in both
early-type and late-type galaxies at similar redshifts and impact
parameters to the supernova sample.  Even though the two samples
cannot be trivially compared, it would be surprising for the supernova
sample to show markedly less extinction than the lens sample.  Second,
from the 4 lenses for which we can derive non-parametric extinction
curves, we find no evidence for gray dust at optical wavelengths.
This eliminates one potential, if physically unlikely, source of
systematic uncertainty for the supernova projects.  Note, however,
that the high $R_V$ extinction curves are quite gray compared to the
typical $R_V=3.1$ curve.  Third, we find additional evidence that
extinction curves are not universal and can deviate significantly from
``standard'' Galactic dust even in regions of low total extinction.
The uncertainty in the appropriate extinction curve must be regarded
as a source of systematic uncertainty.  Since dust grain distributions
depend on the history of dust production and destruction, which in
turn depends heavily on the metallicity and the mean star formation
and energy injection rates, it would be surprising if the ``typical''
extinction curve did not evolve with redshift.

In Figure 3 we compare the total extinction estimates for the bluest
lens images to the raw extinction estimates for the Type Ia supernovae
in Riess et al. (1998).  We could not compare to the Perlmutter et
al. (1997) sample because they did not estimate extinctions.  As with
the estimates of the total extinctions of the gravitational lenses,
the intrinsic colors of the supernovae are too poorly constrained to
allow accurate individual extinction estimates.  The extinction
distribution of the supernovae is peculiar because 11 of the 15
extinction estimates are negative.\footnote{The processed extinctions
appearing in Riess et al. (1998) have been corrected for photometric
errors using a Bayesian procedure which assumes that negative
extinction estimates are due only to photometric errors. The 11
negative extinctions are reset to zero.  Here we show the raw,
uncorrected extinctions (Riess 1998).}  Figure 3 also shows the
expected distribution of extinctions given the uncertainty estimates
for the individual measurements using either the MLCS (typical
$\sigma_E=0.13$ mag) or $\Delta m_{15}$ (typical $\sigma_e=0.10$ mag)
analysis methods.  In both cases, the distribution of the measurements
is significantly broader than the model, and simple statistical tests
show that the distribution of supernova extinctions is inconsistent
with the stated uncertainties and random errors even if all supernovae
had zero intrinsic extinction at a slightly greater than 2-$\sigma$
confidence level.  The best fit to the observed extinction
distribution requires that the true uncertainties in the extinction
be $\sigma_E \simeq 0.18$ mag and that the MLCS and $\Delta m_{15}$
methods underestimate the true uncertainties by a factors of 1.5
and $1.7$ respectively.

There are four possible interpretations of the inconsistency.  First,
it may be genuine sample variance. However, the 2-$\sigma$ confidence
level for the existence of the inconsistency was derived without the
inclusion of the 5 ``snapshot'' supernovae, which also show a
preponderance of negative extinctions.  Second, the Type Ia fitting
methods are underestimating errors by a factor of 1.5--1.7.  The
covariance between extinction and distance means that we must rescale
the distance errors by the same factor as the extinction errors.  The
rescaling has serious cosmological consequences because for 50\%
larger errors, a 2-$\sigma$ confidence region in the
$\Omega_0$--$\Lambda_0$ plane becomes a less than 1-$\sigma$
confidence region.  Third, an estimate of zero extinction in the
supernova sample may not really be zero.  The Type Ia extinction
estimates are actually differential measurements relative to
supernovae in nearby ellipticals which are defined to have zero
extinction.  If these systems have non-zero extinction due to diffuse
dust such as we appear to see in many of the lenses, then the true
zero-point of the supernovae extinction scale might be somewhat
negative.  In this case the existing cosmological limits are biased
towards high $\Omega_0$ (low $\Lambda_0$) by the Bayesian procedure
for renormalizing negative extinctions.  Fourth, there may be a
systematic error or evolutionary effect making the high redshift
supernovae systematically bluer than the low redshift comparison
samples.  Which of these solutions is correct will only become
apparent in larger samples, but it is clear that the derivation of a
physical, self-consistent extinction distribution for the supernovae
is a powerful test of the assumptions and uncertainty estimates used
in the supernova cosmology programs.
   
\section{Summary}

We measured 37 extinctions in 23 gravitational lenses, which
constitutes the largest set of direct extinction measurements outside
the local universe.  We can accurately measure small differential
extinctions ($|\Delta E(B-V)| \gtorder 0.02$ mag) or large absolute
extinctions ($E(B-V) \gtorder 0.1$ mag) because the former depends
only on observed spectral differences while the latter requires a
model for the intrinsic spectrum.  The median differential extinctions
of $\Delta E(B-V)=0.04$ ($0.06$) mag for optically (radio) selected
lenses are modest.  There are two exceptions, both of which are
face-on, gas-rich spiral lenses with differential extinctions of $\Delta E(B-V) =
0.9$ and $3.0$ mag respectively.  We see no pattern of lensed images
at small impact parameters showing systematically higher extinctions
than those at large impact parameters.  If we assume a model for the
intrinsic color of the quasars based on the composite quasar spectra
of Elvis et al. (1994) and Francis et al. (1991) we can estimate the
total extinction.  We find a median total extinction for the bluest
images of $E(B-V) = 0.08$ mag, but the distribution is dominated by
the uncertainties in the intrinsic colors of the quasars and AGN.  The
similarity of the mean and differential extinctions and the lack of a
correlation between image radius and extinction suggests that the
interstellar medium of the lenses is patchy.  For most lenses, the
spectral differences cannot be fully explained by extinction using a
parametric extinction curve and the measured photometric errors --
deviations from the parametric law, systematic underestimates of the
photometric errors, time variability, and microlensing must contribute
to the spectral differences.

Most lens galaxies are early-type galaxies based on their colors,
morphologies, luminosities (Keeton et al. 1998) and they lie on the
fundamental plane (Kochanek et al. 1998a).  Thus, the lens sample
appears to provide unambiguous evidence for patchy, diffuse dust in
some early-type galaxies on scales of $\sim 5h_{65}^{-1}$ kpc (1\farcs0 at
$z=0.5$). There is considerable evidence for dusty disks and patchy
dust in the central regions ($<$ 1 kpc) of local early-type galaxies
(e.g. Schweizer 1987, Kormendy \& Stauffer 1987, Goudfrooij et
al. 1994, van Dokkum \& Franx 1995).  In fact, van Dokkum \& Franx
estimate that 80\% of ellipticals contain nuclear dust once
corrections are made for the difficulty of detecting face-on dusty
disks.  The amount of dust on the larger scales (1--5 kpc) typical of
the impact parameters of the lensed images is less clear in the local
samples, but Goudfrooij \& de Jong (1995) argue for the existence of a
diffuse dust component with $0.04 \ltorder E(B-V) \ltorder 0.25$ to
explain the discrepancy between the IRAS flux presumed to arise from
dust emission and the mass of patchily distributed dust derived from
optical images.  Their estimate of a diffuse dust component is
consistent with the estimate of $E(B-V)=0.10\pm0.08$ mag by Falco et
al. (1998) from comparisons of optical and radio lens surveys.

The two lenses with the highest differential and mean extinctions, the
face-on late-type lenses B~0218+357 and PKS~1830--211, produce
molecular and atomic absorption features in the radio and millimeter
continuum of the quasar (Carilli et al. 1993, Wiklind \& Combes 1995,
1996).  The inferred molecular column densities combined with standard
dust-to-gas ratios overpredict the amount of extinction by factors of
2--5.  No detailed analysis is necessary to realize there is a problem
because the standard dust-to-gas ratios predict $A_V > 20$ mag -- such
large extinctions would make the sources intrinsically more luminous
than any known quasar.  For both heavily extincted images we infer an
extinction curve with $R_V \simeq 6$--$7$ instead of $3.1$, and the
properties of the absorbing regions are very similar to those regions of the
Galaxy where high $R_V$ extinction curves are measured.  The remaining
good candidate for the detection of molecular absorption lines in the
sample is the lensed radio quasar MG~0414+0534 with $z_l=0.96$ and
$z_s=2.64$.  We estimate $N(H_2) \simeq 7 \times 10^{20}$ cm$^{-2}$
based on the largest differential extinction or $N(H_2) \gtorder 4
\times 10^{21}$ cm$^{-2}$ based on the total extinction, although we
expect the total extinction to be associated with the source rather
than the lens.

For systems with sufficient differential extinction, we can estimate
both the extinction law (Nadeau et al. 1991) and the redshift of the
dust (Jean \& Surdej 1998).  Many of the estimated
extinction curves do not match the standard Galactic $R_V=3.1$ curve.
For example, the $z_l=0.96$ elliptical lens MG~0414+0531 has
$R_V=1.5\pm0.2$ and the $z_l=0.68$ spiral lens B~0218+357 has
$R_V=7.2\pm0.1$.  Dust redshifts appear to
work reasonably well provided there is sufficient differential
extinction to determine the extinction curve at a fixed
redshift accurately.  For the systems where the lens redshift is known, the
agreement between the spectroscopic and dust redshifts suggests that
the derived extinction curves are correct -- it is difficult to see
how a systematic error can both mimic a deviant extinction curve and
select out the true lens redshift.  The dust redshifts can be
remarkably accurate, with $\langle z_{dust} - z_l \rangle =
0.00\pm0.08 $, if we confine the comparison to the systems with high
differential extinction ($E(B-V)>0.05$ mag for $R_V=3.1$) and drop 
incorrect secondary redshift solutions. However, when there are
multiple solutions, the systematic uncertainties in the data and the
extinction curve models make it dangerous to choose between the minima
based on simple Gaussian statistics.

The extinctions have a significant effect on determinations of the
cosmological model from the statistics of lensed quasars and the
brightness of Type Ia supernovae. Our directly measured extinction 
distributions are consistent with the
statistical estimates from a comparison of radio-selected and
optically-selected lens surveys by Falco et al. (1998).  {\it Any
cosmological estimate using the statistics of lensed quasars requires
a substantial correction for extinction. }  Both recent
attempts to reconcile the statistics of lensed quasars with the
results of the Type Ia supernovae surveys (Chiba \& Yoshii 1998, Cheng
\& Krauss 1998) neglected the effects of extinction, thereby producing
significant overestimates in the expected number of lensed quasars in
any given cosmology. Since we are unable to determine the distribution of
total extinctions in the gravitational lenses accurately given the
uncertainties in the intrinsic source colors, only the statistics of
lensed radio sources can be used to directly estimate the cosmological
model using gravitational lens statistics (Falco et al. 1998, Cooray
1998).  The statistics of lensed quasars can be used to estimate
the cosmological model only after correcting for the total extinction
derived from a comparison of the radio and optical lens samples.  Extinction
determinations for high redshift Type Ia supernovae (Riess et al. 1998,
Perlmutter et al. 1997) have similar accuracy problems, which makes
it difficult to compare the extinction distributions of the lens
and supernova samples.  We did find, however, an internal inconsistency 
at greater than 2-$\sigma$ confidence in the Riess et al. (1998) supernovae 
extinctions and their uncertainties, whose most obvious symptom is that 
75\% of the supernovae extinctions are negative.  While the origin of the
inconsistency is as yet unclear, the derivation of self-consistent,
physical extinction distributions for the supernovae is a critical
test of the results of the supernova cosmology projects.

The existing lens data were never intended for extinction studies, so
there is enormous potential for improvement.  Systematic errors can be
minimized by obtaining all the necessary multiwavelength photometry at
a common epoch.  Broader wavelength coverage, particularly the
addition of B and U band photometry, would greatly improve the
precision of the estimates for the extinction, the extinction curves,
and lens redshifts.  For many lenses the photometry could be done from
the ground because there will be little contamination from the
early-type lens galaxies at these wavelengths, although HST is
required for the smaller separation systems.  The 2175\AA\ dust
feature should be detectable for many lenses if their extinction laws
resemble that for the Galaxy or the LMC rather than the SMC.  The
feature is redshifted to more easily observed wavelengths, appearing
at $2175(1+z_l)$\AA.  Although observable from the ground for lenses
with redshifts $z_l \gtorder 0.8$, precision spectrophotometry of the
closely spaced images or lower redshift lenses will require HST/STIS
to detect or limit the presence of the feature.

\acknowledgements We would like to thank A. Goodman and M. Spaans for
discussions of extinction, and A. Riess and P. Garnavich for
discussions of the Type Ia extinction estimates.  Support for the
CASTLES project was provided by NASA through grant numbers GO-7495 and
GO-7887 from the Space Telescope Science Institute, which is operated
by the Association of Universities for Research in Astronomy, Inc.
CSK and CRK were also supported by the NASA Astrophysics Theory
Program grant NAG5-4062.  HWR is also supported by a Fellowship from
the Alfred P. Sloan Foundation.  This research was supported in part
by the Smithsonian Institution.

\vfill\eject

\def\mult1c#1{\multicolumn{1}{c}{#1}}
\begin{deluxetable}{lcrrrrl}
\scriptsize
\tablecaption{Magnitude Differences }
\tablewidth{0pt}
\tablehead{\mult1c{Lens} &$\lambda^{-1}$ &\mult1c{Image 1} &\mult1c{Image 2} 
                                         &\mult1c{Image 3} &\mult1c{Image 4} &Source \\
   &$\mu$m$^{-1}$  &\mult1c{mag} &\mult1c{mag}  
                   &\mult1c{mag} &\mult1c{mag}     & }
\startdata
\input tab1.tex
\enddata
\tablecomments{Magnitude differences relative to image \#1 as a
function of the inverse central wavelength of the filter.  Radio flux
ratios have been converted to magnitude differences.  We use the
``standard order'' for the images in the literature which are labeled
alphabetically or with some other system. Thus for images labeled A--D
the correspondence is \#1=A, \#2=B, \#3=C, \#4=D, or for systems with
close pairs discovered at a later date \#1=A1, \#2=A2, \#3=B, \#4=D, 
etc. }
\end{deluxetable}

\vfill\eject
\begin{deluxetable}{lcccccccc}
\small
\tablecaption{Dust Statistics}
\tablewidth{0pt}
\tablehead{\mult1c{Lens} &Code &$z_{dust}$ &$\chi^2_1/N_{dof}$ &$\sigma_{sys}$ &$\chi^2_2/N_{dof}$ &$\sigma_{sys}$ &$\chi^2_3/N_{dof}$ &$R_V$ \\ 
                         &     &           &no dust            &mag            &$R_V\equiv3.1$     &mag            &variable $R_V$     &      }
\startdata
\input tab2.tex
\enddata

\tablecomments{ The codes are R for radio-selected, O for
  optically-selected, E for early-type lenses, L for late-type lenses,
  and ? for untyped lenses.  Since the lens population is known to be
  dominated by early-type galaxies, we treat the untyped lenses as
  early-type galaxies when examining subsamples.  The lens redshift is
  in parentheses if it is an estimated redshift rather than a
  spectroscopically measured redshift.  We quote the goodness of fit
  in the form $\chi^2/N_{dof}$ where $\chi^2$ is the value of the
  statistic, $N_{dof}$ is the number of degrees of freedom, and a good
  fit would have $\chi^2 = N_{dof} \pm (2N_{dof})^{1/2}$ for $N_{dof}
  \gg 1$.  The value of $\sigma_{sys}$ provides an estimate for the
  systematic error in magnitudes beyond the estimated photometric
  errors needed to make $\chi^2 = N_{dof}$. }
\end{deluxetable}

\vfill\eject
\begin{deluxetable}{lcccccr}
\footnotesize
\tablecaption{Differential and Total Extinctions ($R_V\equiv3.1$) }
\tablewidth{0pt}
\tablehead{\mult1c{Lens} &Image 1            &Image 2          &Image 3         &Image 4         &$A_V$  &$E(B-V)$ \\
                         &$\Delta E(B-V)$    &$\Delta E(B-V)$  &$\Delta E(B-V)$ &$\Delta E(B-V)$ &max    &\mult1c{total}      }
\startdata
\input tab3.tex
\enddata
\tablecomments{All differential extinctions are presented relative to
  the least reddened (bluest) image. The uncertainties have been
  broadened when $\chi_2^2 > N_{dof}$ by the factor
  $(\chi_2^2/N_{dof})^{1/2}$ to compensate for possible underestimates
  of the photometric errors.  This is equivalent to including the
  effects of $\sigma_{sys}$.  The maximum extinction $A_V$ is
  determined by requiring $M_B > -29$ mag ($H_0=50$
  km~s$^{-1}$~Mpc$^{-1}$) corrected for a fixed lens magnification of
  10.  The error bars in the total extinction of the bluest image
  assume the intrinsic color is uncertain by 0.1 mag. We include no
  estimates for the total extinctions of HST~12531--2914 and
  HST~14176+5226 because the sources are probably not quasars. }
\end{deluxetable}

\vfill\eject
\begin{deluxetable}{lccc}
\small
\tablecaption{Extinction Corrected Flux Ratios ($R_V\equiv3.1$) }
\tablewidth{0pt}
\tablehead{Lens &Image 2  &Image 3 &Image 4 }
\startdata
\input tab4.tex
\enddata
\tablecomments{ The flux ratios are calculated relative to Image 1
  which is defined to have a flux of unity.  The uncertainties in the
  true flux ratios have been broadened when $\chi_2^2 > N_{dof}$ by
  the factor $(\chi_2^2/N_{dof})^{1/2}$ to compensate for possible
  underestimates of the photometric errors.  This is equivalent to
  including the effects of $\sigma_{sys}$.  The extinction law is
  nearly self-similar in the optical and infrared, so an incorrect
  lens redshift creates much larger errors in $E(B-V)$ and
  $R(\lambda)$ than in their product ($R(\lambda)E(B-V)$), and only
  the product is needed for estimates of the intrinsic flux ratios. 
  We do not include HST~12531--2914 and HST~14176+5226 because their
  intrinsic flux ratios cannot be estimated with any accuracy if
  extinction estimates are required.} 
\end{deluxetable}

\vfill\eject
\begin{deluxetable}{lccccl}
\small
\tablecaption{Dust Redshifts}
\tablewidth{0pt}
\tablehead{Lens &Redshift &$z_{dust}$  &low  &high  &Comments }
\startdata
 B0218+357      &0.68     &0.59      &0.57      &0.59  &Formal uncertainty too small \\
 MG0414+0534    &0.96     &0.23      &0.15      &0.29  & \\
                &         &1.08      &1.01      &1.13  &Formal $\Delta\chi^2=7.1$ from best fit \\
 FBQ0951+2635   &         &0.30      &0.26      &0.42  & \\
 LBQS1009--0252 &         &0.35      &0.21      &0.44  & \\
                &         &0.86      &0.00      &1.50  &Formal $\Delta\chi^2=4.6$ from best fit \\
 H1413+117      &         &0.34      &0.30      &0.44  & \\
 B1422+231      &0.34     &0.72      &0.60      &0.80  &DUST REDSHIFT WRONG \\
 B1600+434      &0.42     &0.27      &0.12      &0.41  & \\
                &         &0.83      &0.74      &1.50  &Formal $\Delta\chi^2=2.5$ from best fit \\
 MG2016+112     &1.01     &1.05      &0.89      &1.22  & \\
                &         &0.23      &0.08      &0.38  &Formal $\Delta\chi^2=1.2$ from best fit \\
 Q2237+0305     &0.04     &0.00      &0.00      &0.00  &Formal uncertainty too small \\
\enddata
\tablecomments{ Dust redshift estimates for systems with $\Delta
   E(B-V) > 0.05$ mag and standard errors $\sigma(\Delta E(B-V)) <
   0.1$ for $R_V=3.1$ plus B~1422+231.  We added a Gaussian prior that
   $R_V=3.1\pm1.0$, and present the standard error in the redshift
   about the minimum (where $\Delta\chi^2(z_{dust})=1$).  We present
   all systems where the dust redshift was measured (meaning that the
   redshift uncertainty at the best fit redshift had a full width
   smaller than $\Delta z_{dust} < 0.5$).  The lenses SBS~0909+532,
   HE~1104-1805, and SBS~1520+530 were dropped because $\Delta
   z_{dust} > 0.5$.  For the secondary minima, the standard errors are
   computed from the value of $\chi^2$ at the secondary minimum.  The
   resulting redshift uncertainties sometimes encompass the primary
   minimum (e.g. LBQS1009--0252).  }
\end{deluxetable}

\end{document}

%% file: tab1.tex
               &$ 1.48 $&$\equiv 0\pm 0.05 $&$  2.35 \pm  0.03 $&$                 $&$                 $ &Leh\'ar et al. (1999)                     \\
               &$ 0.63 $&$\equiv 0\pm 0.02 $&$  2.29 \pm  0.02 $&$                 $&$                 $ &Leh\'ar et al. (1999)                     \\
B0218+357      &$ 2.92 $&$\equiv 0\pm 0.20 $&$ -1.10 \pm  0.10 $&$                 $&$                 $ &Leh\'ar et al. (1999)                     \\
               &$ 1.80 $&$\equiv 0\pm 0.10 $&$ -2.25 \pm  0.03 $&$                 $&$                 $ &Leh\'ar et al. (1999)                     \\
               &$ 1.23 $&$\equiv 0\pm 0.01 $&$ -2.40 \pm  0.01 $&$                 $&$                 $ &Leh\'ar et al. (1999)                     \\
               &$ 0.63 $&$\equiv 0\pm 0.01 $&$ -0.57 \pm  0.01 $&$                 $&$                 $ &Leh\'ar et al. (1999)                     \\
               &$ 0.00 $&$\equiv 0\pm 0.01 $&$  1.40 \pm  0.02 $&$                 $&$                 $ &Patnaik et al. (1995)                   \\
MG0414+0534    &$ 1.48 $&$\equiv 0\pm 0.03 $&$  1.18 \pm  0.03 $&$  0.73 \pm  0.03 $&$  1.56 \pm  0.04 $ &CASTLES                                 \\
               &$ 1.23 $&$\equiv 0\pm 0.02 $&$  0.94 \pm  0.02 $&$  0.81 \pm  0.02 $&$  1.67 \pm  0.06 $ &CASTLES                                 \\
               &$ 0.91 $&$\equiv 0\pm 0.02 $&$  0.49 \pm  0.02 $&$  0.95 \pm  0.02 $&$  1.78 \pm  0.02 $ &CASTLES                                 \\
               &$ 0.63 $&$\equiv 0\pm 0.03 $&$  0.32 \pm  0.03 $&$  1.01 \pm  0.03 $&$  1.86 \pm  0.03 $ &CASTLES                                 \\
               &$ 0.49 $&$\equiv 0\pm 0.03 $&$  0.20 \pm  0.03 $&$  1.04 \pm  0.03 $&$  1.90 \pm  0.03 $ &CASTLES                                 \\
               &$ 0.00 $&$\equiv 0\pm 0.04 $&$  0.13 \pm  0.04 $&$  1.04 \pm  0.04 $&$  2.09 \pm  0.04 $ &Moore \& Hewitt (1997)                  \\
SBS0909+532    &$ 2.25 $&$\equiv 0\pm 0.04 $&$  1.29 \pm  0.04 $&$                 $&$                 $ &Kochanek et al. (1997)                  \\
               &$ 1.52 $&$\equiv 0\pm 0.04 $&$  0.58 \pm  0.04 $&$                 $&$                 $ &Kochanek et al. (1997)                  \\
               &$ 1.24 $&$\equiv 0\pm 0.04 $&$  0.31 \pm  0.04 $&$                 $&$                 $ &Kochanek et al. (1997)                  \\
               &$ 0.63 $&$\equiv 0\pm 0.01 $&$  0.13 \pm  0.03 $&$                 $&$                 $ &Leh\'ar et al. (1999)                     \\
FBQ0951+2635   &$ 2.24 $&$\equiv 0\pm 0.03 $&$  0.89 \pm  0.03 $&$                 $&$                 $ &Schechter et al. (1998)                 \\
               &$ 1.80 $&$\equiv 0\pm 0.03 $&$  0.94 \pm  0.03 $&$                 $&$                 $ &Schechter et al. (1998)                 \\
               &$ 1.52 $&$\equiv 0\pm 0.03 $&$  0.96 \pm  0.03 $&$                 $&$                 $ &Schechter et al. (1998)                 \\
               &$ 1.24 $&$\equiv 0\pm 0.03 $&$  0.95 \pm  0.03 $&$                 $&$                 $ &Schechter et al. (1998)                 \\
               &$ 0.63 $&$\equiv 0\pm 0.02 $&$  1.36 \pm  0.02 $&$                 $&$                 $ &CASTLES                                 \\
               &$ 0.00 $&$\equiv 0\pm 0.04 $&$  1.67 \pm  0.18 $&$                 $&$                 $ &Schechter et al. (1998)                 \\
BRI0952-0115   &$ 1.49 $&$\equiv 0\pm 0.04 $&$  1.29 \pm  0.03 $&$                 $&$                 $ &Leh\'ar et al. (1999)                     \\
               &$ 0.63 $&$\equiv 0\pm 0.01 $&$  1.37 \pm  0.01 $&$                 $&$                 $ &Leh\'ar et al. (1999)                     \\
Q0957+561      &$ 4.03 $&$\equiv 0\pm 0.05 $&$  0.06 \pm  0.05 $&$                 $&$                 $ &Dolan et al. (1995)                     \\
               &$ 3.52 $&$\equiv 0\pm 0.05 $&$  0.06 \pm  0.05 $&$                 $&$                 $ &Dolan et al. (1995)                     \\
               &$ 1.80 $&$\equiv 0\pm 0.01 $&$  0.01 \pm  0.02 $&$                 $&$                 $ &Bernstein et al. (1997)                 \\
               &$ 1.23 $&$\equiv 0\pm 0.02 $&$  0.02 \pm  0.02 $&$                 $&$                 $ &Bernstein et al. (1997)                 \\
               &$ 0.63 $&$\equiv 0\pm 0.01 $&$  0.08 \pm  0.01 $&$                 $&$                 $ &CASTLES                                 \\
               &$ 0.00 $&$\equiv 0\pm 0.02 $&$  0.32 \pm  0.02 $&$                 $&$                 $ &Conner et al. (1992)                    \\
LBQS1009-0252  &$ 2.25 $&$\equiv 0\pm 0.14 $&$  2.52 \pm  0.14 $&$                 $&$                 $ &Hewett et al. (1994)                    \\
               &$ 1.82 $&$\equiv 0\pm 0.07 $&$  2.62 \pm  0.07 $&$                 $&$                 $ &Hewett et al. (1994)                    \\
               &$ 1.50 $&$\equiv 0\pm 0.07 $&$  2.26 \pm  0.07 $&$                 $&$                 $ &Hewett et al. (1994)                    \\
               &$ 1.25 $&$\equiv 0\pm 0.06 $&$  2.03 \pm  0.06 $&$                 $&$                 $ &Hewett et al. (1994)                    \\
               &$ 0.63 $&$\equiv 0\pm 0.02 $&$  1.58 \pm  0.03 $&$                 $&$                 $ &Leh\'ar et al. (1999)                     \\
      \hline
 \tablebreak
Q1017-207=J03  &$ 1.80 $&$\equiv 0\pm 0.02 $&$  2.16 \pm  0.03 $&$                 $&$                 $ &Leh\'ar et al. (1999)                     \\
               &$ 1.23 $&$\equiv 0\pm 0.02 $&$  2.15 \pm  0.02 $&$                 $&$                 $ &Leh\'ar et al. (1999)                     \\
               &$ 0.63 $&$\equiv 0\pm 0.03 $&$  2.15 \pm  0.04 $&$                 $&$                 $ &Leh\'ar et al. (1999)                     \\
B1030+074      &$ 1.80 $&$\equiv 0\pm 0.01 $&$  3.08 \pm  0.04 $&$                 $&$                 $ &Leh\'ar et al. (1999)                     \\
               &$ 1.23 $&$\equiv 0\pm 0.02 $&$  3.05 \pm  0.09 $&$                 $&$                 $ &Leh\'ar et al. (1999)                     \\
               &$ 0.00 $&$\equiv 0\pm 0.17 $&$  2.96 \pm  0.17 $&$                 $&\multicolumn{2}{r}{Xanthopoulos et al. (1998)} \\
HE1104-1805    &$ 1.80 $&$\equiv 0\pm 0.01 $&$  1.76 \pm  0.03 $&$                 $&$                 $ &Leh\'ar et al. (1999)                     \\
               &$ 1.23 $&$\equiv 0\pm 0.01 $&$  1.61 \pm  0.02 $&$                 $&$                 $ &Leh\'ar et al. (1999)                     \\
               &$ 0.63 $&$\equiv 0\pm 0.03 $&$  1.47 \pm  0.03 $&$                 $&$                 $ &Leh\'ar et al. (1999)                     \\
PG1115+080     &$ 1.80 $&$\equiv 0\pm 0.01 $&$  0.45 \pm  0.01 $&$  1.97 \pm  0.01 $&$  1.47 \pm  0.01 $ &CASTLES                                 \\
               &$ 1.23 $&$\equiv 0\pm 0.02 $&$  0.39 \pm  0.01 $&$  1.96 \pm  0.01 $&$  1.46 \pm  0.02 $ &CASTLES                                 \\
               &$ 0.63 $&$\equiv 0\pm 0.03 $&$  0.48 \pm  0.03 $&$  1.93 \pm  0.02 $&$  1.48 \pm  0.05 $ &Impey et al. (1998)                     \\
Q1208+1011     &$ 2.28 $&$\equiv 0\pm 0.04 $&$  1.36 \pm  0.04 $&$                 $&$                 $ &Bahcall et al. (1992)                   \\
               &$ 1.80 $&$\equiv 0\pm 0.02 $&$  1.56 \pm  0.02 $&$                 $&$                 $ &Bahcall et al. (1992)                   \\
               &$ 1.27 $&$\equiv 0\pm 0.02 $&$  1.51 \pm  0.02 $&$                 $&$                 $ &Bahcall et al. (1992)                   \\
               &$ 1.42 $&$\equiv 0\pm 0.02 $&$  1.51 \pm  0.02 $&$                 $&$                 $ &Bahcall et al. (1992)                   \\
               &$ 0.63 $&$\equiv 0\pm 0.01 $&$  1.63 \pm  0.01 $&$                 $&$                 $ &Leh\'ar et al. (1999)                     \\
HST12531-2914  &$ 1.65 $&$\equiv 0\pm 0.15 $&$ -0.13 \pm  0.15 $&$ -0.30 \pm  0.11 $&$  0.49 \pm  0.24 $ &Ratnatunga et al. (1995)                \\
               &$ 1.23 $&$\equiv 0\pm 0.24 $&$ -0.31 \pm  0.17 $&$ -0.39 \pm  0.18 $&$ -0.08 \pm  0.24 $ &Ratnatunga et al. (1995)                \\
H1413+117      &$ 2.98 $&$\equiv 0\pm 0.03 $&$  0.34 \pm  0.02 $&$ -0.02 \pm  0.02 $&$  0.17 \pm  0.03 $ &Turnshek et al. (1997)                  \\
               &$ 2.97 $&$\equiv 0\pm 0.02 $&$  0.28 \pm  0.02 $&$ -0.07 \pm  0.02 $&$  0.21 \pm  0.02 $ &Monier et al. (1998)                    \\
               &$ 2.67 $&$\equiv 0\pm 0.02 $&$  0.26 \pm  0.02 $&$  0.05 \pm  0.02 $&$  0.17 \pm  0.02 $ &Monier et al. (1998)                    \\
               &$ 2.35 $&$\equiv 0\pm 0.02 $&$  0.23 \pm  0.02 $&$  0.09 \pm  0.02 $&$  0.40 \pm  0.02 $ &Monier et al. (1998)                    \\
               &$ 2.12 $&$\equiv 0\pm 0.02 $&$  0.21 \pm  0.02 $&$  0.20 \pm  0.02 $&$  0.21 \pm  0.02 $ &Monier et al. (1998)                    \\
               &$ 1.90 $&$\equiv 0\pm 0.02 $&$  0.22 \pm  0.02 $&$  0.20 \pm  0.02 $&$  0.33 \pm  0.02 $ &Monier et al. (1998)                    \\
               &$ 1.81 $&$\equiv 0\pm 0.05 $&$  0.17 \pm  0.05 $&$  0.23 \pm  0.05 $&$  0.31 \pm  0.05 $ &Turnshek et al. (1997)                  \\
               &$ 1.81 $&$\equiv 0\pm 0.04 $&$  0.23 \pm  0.04 $&$  0.25 \pm  0.04 $&$  0.43 \pm  0.04 $ &Turnshek et al. (1997)                  \\
               &$ 1.81 $&$\equiv 0\pm 0.09 $&$  0.31 \pm  0.05 $&$  0.35 \pm  0.06 $&$  0.49 \pm  0.06 $ &Turnshek et al. (1997)                  \\
               &$ 1.74 $&$\equiv 0\pm 0.02 $&$  0.20 \pm  0.02 $&$  0.23 \pm  0.02 $&$  0.34 \pm  0.02 $ &Monier et al. (1998)                    \\
               &$ 1.60 $&$\equiv 0\pm 0.02 $&$  0.20 \pm  0.02 $&$  0.22 \pm  0.02 $&$  0.31 \pm  0.02 $ &Monier et al. (1998)                    \\
               &$ 1.42 $&$\equiv 0\pm 0.02 $&$  0.17 \pm  0.03 $&$  0.31 \pm  0.03 $&$  0.41 \pm  0.02 $ &Turnshek et al. (1997)                  \\
               &$ 1.42 $&$\equiv 0\pm 0.02 $&$  0.15 \pm  0.02 $&$  0.28 \pm  0.02 $&$  0.35 \pm  0.02 $ &Turnshek et al. (1997)                  \\
               &$ 1.23 $&$\equiv 0\pm 0.02 $&$  0.10 \pm  0.02 $&$  0.30 \pm  0.02 $&$  0.40 \pm  0.02 $ &Turnshek et al. (1997)                  \\
               &$ 0.63 $&$\equiv 0\pm 0.01 $&$  0.10 \pm  0.01 $&$  0.38 \pm  0.01 $&$  0.62 \pm  0.02 $ &McLeod et al. (1998)                    \\
               &$ 0.00 $&$\equiv 0\pm 0.10 $&$  0.20 \pm  0.10 $&$  0.61 \pm  0.10 $&$  0.53 \pm  0.10 $ &McLeod et al. (1998)                    \\
      \hline
 \tablebreak
HST14176+5226  &$ 1.65 $&$\equiv 0\pm 0.06 $&$  0.14 \pm  0.07 $&$  0.36 \pm  0.08 $&$  0.34 \pm  0.08 $ &Ratnatunga et al. (1995)                \\
               &$ 1.23 $&$\equiv 0\pm 0.08 $&$  0.26 \pm  0.09 $&$  0.35 \pm  0.10 $&$  0.43 \pm  0.12 $ &Ratnatunga et al. (1995)                \\
B1422+231      &$ 2.92 $&$\equiv 0\pm 0.03 $&$ -0.42 \pm  0.03 $&$  0.26 \pm  0.03 $&$  3.62 \pm  0.03 $ &Impey et al. (1996)                     \\
               &$ 2.09 $&$\equiv 0\pm 0.02 $&$ -0.28 \pm  0.02 $&$  0.50 \pm  0.02 $&$  3.68 \pm  0.05 $ &Impey et al. (1996)                     \\
               &$ 2.08 $&$\equiv 0\pm 0.03 $&$ -0.25 \pm  0.03 $&$  0.41 \pm  0.03 $&$  3.81 \pm  0.03 $ &Impey et al. (1996)                     \\
               &$ 2.03 $&$\equiv 0\pm 0.03 $&$ -0.28 \pm  0.03 $&$  0.52 \pm  0.03 $&$  3.64 \pm  0.03 $ &Yee \& Ellingson (1994)                 \\
               &$ 1.90 $&$\equiv 0\pm 0.02 $&$ -0.29 \pm  0.02 $&$  0.50 \pm  0.02 $&$  3.72 \pm  0.05 $ &Impey et al. (1996)                     \\
               &$ 1.82 $&$\equiv 0\pm 0.05 $&$ -0.25 \pm  0.05 $&$  0.55 \pm  0.05 $&$  3.71 \pm  0.10 $ &Remy et al. (1993)                      \\
               &$ 1.74 $&$\equiv 0\pm 0.02 $&$ -0.20 \pm  0.02 $&$  0.57 \pm  0.02 $&$  3.76 \pm  0.05 $ &Impey et al. (1996)                     \\
               &$ 1.60 $&$\equiv 0\pm 0.02 $&$ -0.26 \pm  0.02 $&$  0.50 \pm  0.02 $&$  3.65 \pm  0.04 $ &Impey et al. (1996)                     \\
               &$ 1.52 $&$\equiv 0\pm 0.05 $&$ -0.33 \pm  0.05 $&$  0.47 \pm  0.05 $&$  3.13 \pm  0.10 $ &Remy et al. (1993)                      \\
               &$ 1.25 $&$\equiv 0\pm 0.05 $&$ -0.29 \pm  0.05 $&$  0.49 \pm  0.05 $&$  3.16 \pm  0.10 $ &Remy et al. (1993)                      \\
               &$ 1.53 $&$\equiv 0\pm 0.03 $&$ -0.32 \pm  0.03 $&$  0.48 \pm  0.03 $&$  3.63 \pm  0.03 $ &Yee \& Ellingson (1994)                 \\
               &$ 0.63 $&$\equiv 0\pm 0.01 $&$ -0.12 \pm  0.02 $&$  0.56 \pm  0.02 $&$  3.73 \pm  0.02 $ &CASTLES                                 \\
               &$ 0.00 $&$\equiv 0\pm 0.03 $&$ -0.03 \pm  0.03 $&$  0.69 \pm  0.03 $&$  4.20 \pm  0.12 $ &Patnaik et al. (1992)                   \\
SBS1520+530    &$ 2.25 $&$\equiv 0\pm 0.04 $&$  0.28 \pm  0.08 $&$                 $&$                 $ &Chavushyan et al. (1997)                \\
               &$ 1.80 $&$\equiv 0\pm 0.04 $&$  0.53 \pm  0.02 $&$                 $&$                 $ &Chavushyan et al. (1997)                \\
               &$ 1.44 $&$\equiv 0\pm 0.05 $&$  0.65 \pm  0.05 $&$                 $&$                 $ &Chavushyan et al. (1997)                \\
               &$ 1.14 $&$\equiv 0\pm 0.11 $&$  0.58 \pm  0.15 $&$                 $&$                 $ &Chavushyan et al. (1997)                \\
               &$ 0.63 $&$\equiv 0\pm 0.01 $&$  0.83 \pm  0.01 $&$                 $&$                 $ &CASTLES                                 \\
B1600+434      &$ 1.80 $&$\equiv 0\pm 0.07 $&$  0.77 \pm  0.05 $&$                 $&$                 $ &CASTLES                                 \\
               &$ 1.49 $&$\equiv 0\pm 0.02 $&$  0.44 \pm  0.02 $&$                 $&$                 $ &CASTLES                                 \\
               &$ 0.63 $&$\equiv 0\pm 0.02 $&$  0.15 \pm  0.02 $&$                 $&$                 $ &CASTLES                                 \\
               &$ 0.00 $&$\equiv 0\pm 0.05 $&$  0.29 \pm  0.05 $&$                 $&$                 $ &Jackson et al. (1995)                   \\
PKS1830-211    &$ 0.63 $&$\equiv 0\pm 0.01 $&$  5.03 \pm  0.10 $&$                 $&$                 $ &Leh\'ar et al. (1999)                     \\
               &$ 0.49 $&$\equiv 0\pm 0.02 $&$  3.88 \pm  0.03 $&$                 $&$                 $ &Leh\'ar et al. (1999)                     \\
               &$ 0.00 $&$\equiv 0\pm 0.03 $&$  0.46 \pm  0.03 $&$                 $&$                 $ &Lovell et al. (1998)                    \\
MG2016+112     &$ 1.91 $&$\equiv 0\pm 0.03 $&$  0.37 \pm  0.03 $&$                 $&$                 $ &Schneider et al. (1986)                 \\
               &$ 1.91 $&$\equiv 0\pm 0.05 $&$  0.57 \pm  0.05 $&$                 $&$                 $ &Schneider et al. (1985)                 \\
               &$ 1.50 $&$\equiv 0\pm 0.05 $&$  0.16 \pm  0.05 $&$                 $&$                 $ &Schneider et al. (1985)                 \\
               &$ 1.50 $&$\equiv 0\pm 0.10 $&$ -0.07 \pm  0.10 $&$                 $&$                 $ &Schneider et al. (1985)                 \\
               &$ 1.25 $&$\equiv 0\pm 0.05 $&$  0.31 \pm  0.05 $&$                 $&$                 $ &Schneider et al. (1985)                 \\
               &$ 0.63 $&$\equiv 0\pm 0.03 $&$  0.02 \pm  0.02 $&$                 $&$                 $ &CASTLES                                 \\
               &$ 0.00 $&$\equiv 0\pm 0.03 $&$ -0.06 \pm  0.03 $&$                 $&$                 $ &Lawrence et al. (1993)                  \\
      \hline
 \tablebreak
HE2149-2745    &$ 2.25 $&$\equiv 0\pm 0.03 $&$  1.62 \pm  0.03 $&$                 $&$                 $ &Wisotzki et al. (1996)                  \\
               &$ 1.52 $&$\equiv 0\pm 0.03 $&$  1.57 \pm  0.03 $&$                 $&$                 $ &Wisotzki et al. (1996)                  \\
               &$ 0.63 $&$\equiv 0\pm 0.02 $&$  1.57 \pm  0.03 $&$                 $&$                 $ &CASTLES                                 \\
Q2237+0305     &$ 3.33 $&$\equiv 0\pm 0.03 $&$  0.25 \pm  0.03 $&$  1.39 \pm  0.04 $&$  1.56 \pm  0.03 $ &Blanton et al. (1998)                   \\
               &$ 2.98 $&$\equiv 0\pm 0.04 $&$ -0.10 \pm  0.04 $&$  0.93 \pm  0.04 $&$  1.34 \pm  0.04 $ &Rix et al. (1992)                       \\
               &$ 2.98 $&$\equiv 0\pm 0.02 $&$  0.28 \pm  0.04 $&$  1.30 \pm  0.03 $&$  1.45 \pm  0.03 $ &Blanton et al. (1998)                   \\
               &$ 1.82 $&$\equiv 0\pm 0.27 $&$ -0.02 \pm  0.17 $&$  1.02 \pm  0.06 $&$  1.25 \pm  0.08 $ &Ostensen et al. (1996)                  \\
               &$ 1.80 $&$\equiv 0\pm 0.02 $&$  0.03 \pm  0.02 $&$  1.19 \pm  0.03 $&$  1.52 \pm  0.03 $ &CASTLES                                 \\
               &$ 1.44 $&$\equiv 0\pm 0.19 $&$ -0.05 \pm  0.10 $&$  0.84 \pm  0.05 $&$  1.07 \pm  0.06 $ &Ostensen et al. (1996)                  \\
               &$ 1.42 $&$\equiv 0\pm 0.04 $&$ -0.13 \pm  0.04 $&$  0.69 \pm  0.04 $&$  0.92 \pm  0.04 $ &Rix et al. (1992)                       \\
               &$ 1.14 $&$\equiv 0\pm 0.17 $&$ -0.06 \pm  0.08 $&$  0.73 \pm  0.09 $&$  0.97 \pm  0.09 $ &Ostensen et al. (1996)                  \\
               &$ 0.63 $&$\equiv 0\pm 0.03 $&$  0.46 \pm  0.02 $&$  0.69 \pm  0.02 $&$  0.96 \pm  0.03 $ &CASTLES                                 \\
               &$ 0.49 $&$\equiv 0\pm 0.02 $&$  0.53 \pm  0.02 $&$  0.78 \pm  0.02 $&$  1.00 \pm  0.02 $ &CASTLES                                 \\
               &$ 0.00 $&$\equiv 0\pm 0.19 $&$ -0.09 \pm  0.18 $&$  0.65 \pm  0.35 $&$  0.28 \pm  0.25 $ &Falco et al. (1996)                     \\

%% file: tab2.tex
Q0142-100     &OE &$ 0.49$   &$    0.86/  3$ &$        $ &$    0.39/  2$ &$        $ &$    0.39/  1$ &$  3.11 \pm   1.00 $\\
B0218+357     &RL &$ 0.68$   &$23163.40/  5$ &$    1.07$ &$ 2043.45/  4$ &$    2.99$ &$  465.02/  3$ &$  7.20 \pm   0.08 $\\
MG0414+0534   &RE &$ 0.96$   &$ 1239.86/ 16$ &$    0.27$ &$   89.96/ 13$ &$    0.07$ &$   22.61/ 12$ &$  1.47 \pm   0.15 $\\
SBS0909+532   &O? &$ (0.60)$ &$  330.14/  4$ &$    0.34$ &$   26.05/  3$ &$    0.11$ &$    6.19/  2$ &$  0.64 \pm   0.15 $\\
FBQ0951+2635  &OE &$ (0.30)$ &$  155.99/  6$ &$    0.21$ &$   20.72/  5$ &$    0.06$ &$   14.18/  4$ &$  4.86 \pm   0.85 $\\
BRI0952-0115  &OE &$ (0.55)$ &$    2.37/  2$ &$    0.01$ &$    0.00/  1$ &$        $ &$    0.00/  0$ &$  3.10 \pm   1.00 $\\
Q0957+561     &RE &$ 0.36$   &$   85.86/  6$ &$    0.07$ &$   63.48/  5$ &$    0.08$ &$   56.64/  4$ &$  6.63 \pm   0.87 $\\
LBQS1009-0252 &OE &$ (0.60)$ &$  146.16/  5$ &$    0.37$ &$    6.25/  4$ &$    0.06$ &$    6.06/  3$ &$  2.72 \pm   0.84 $\\
Q1017-207=J03 &O? &$ (0.60)$ &$    0.05/  3$ &$        $ &$    0.02/  2$ &$        $ &$    0.02/  1$ &$  3.10 \pm   1.00 $\\
B1030+074     &RE &$ 0.60$   &$    0.31/  3$ &$        $ &$    0.00/  2$ &$        $ &$    0.00/  1$ &$  3.10 \pm   1.00 $\\
HE1104-1805   &OE &$ (0.80)$ &$   31.92/  3$ &$    0.09$ &$    0.82/  2$ &$        $ &$    0.76/  1$ &$  2.87 \pm   0.96 $\\
PG1115+080    &OE &$ 0.31$   &$   15.65/  7$ &$    0.02$ &$   13.46/  4$ &$    0.03$ &$   13.43/  3$ &$  2.89 \pm   0.99 $\\
Q1208+1011    &O? &$ (0.60)$ &$   37.74/  5$ &$    0.07$ &$    8.40/  4$ &$    0.02$ &$    8.25/  3$ &$  3.47 \pm   0.97 $\\
HST12531-2914 &OE &$ (0.81)$ &$    1.89/  4$ &$        $ &$    0.00/  1$ &$        $ &$    0.00/  0$ &$  3.10 \pm   1.00 $\\
H1413+117     &O? &$ (0.70)$ &$ 1235.19/ 49$ &$    0.10$ &$  219.01/ 46$ &$    0.05$ &$  218.96/ 45$ &$  2.94 \pm   0.66 $\\
HST14176+5226 &OE &$ 0.81$   &$    0.95/  4$ &$        $ &$    0.00/  1$ &$        $ &$    0.00/  0$ &$  3.10 \pm   1.00 $\\
B1422+231     &RE &$ 0.34$   &$  206.65/ 37$ &$    0.11$ &$  124.46/ 34$ &$    0.10$ &$  124.41/ 33$ &$  2.91 \pm   0.81 $\\
SBS1520+530   &OE &$ (0.49)$ &$   79.37/  5$ &$    0.21$ &$    3.03/  4$ &$        $ &$    2.95/  3$ &$  2.83 \pm   0.96 $\\
B1600+434     &RL &$ 0.42$   &$   79.99/  4$ &$    0.17$ &$   17.18/  3$ &$    0.11$ &$   10.73/  2$ &$  0.92 \pm   0.58 $\\
PKS1830-211   &RL &$ 0.89$   &$ 4400.75/  4$ &$    1.58$ &$   26.12/  3$ &$    0.24$ &$   10.76/  2$ &$  6.34 \pm   0.16 $\\
MG2016+112    &RE &$ 1.01$   &$  116.83/  9$ &$    0.15$ &$   25.39/  8$ &$    0.10$ &$   25.25/  7$ &$  2.80 \pm   0.78 $\\
HE2149-2745   &OE &$ (0.49)$ &$    0.98/  3$ &$        $ &$    0.35/  2$ &$        $ &$    0.35/  1$ &$  3.08 \pm   1.00 $\\
Q2237+0305    &OL &$ 0.04$   &$ 1459.44/ 32$ &$    0.21$ &$  482.57/ 29$ &$    0.15$ &$  385.38/ 28$ &$  5.29 \pm   0.82 $\\

%% file: tab3.tex
Q0142-100     &$ \equiv 0$       &$ 0.01 \pm  0.02 $&                  &                  &$ 1.3 $&$-0.06 \pm 0.09 $\\
B0218+357     &$ 0.90 \pm  0.14 $&$ \equiv 0$       &                  &                  &$ 3.5 $&$ 0.62 \pm 0.04 $\\
MG0414+0534   &$ 0.09 \pm  0.03 $&$ 0.31 \pm  0.03 $&$ 0.02 \pm  0.03 $&$ \equiv 0$       &$ 5.7 $&$ 1.41 \pm 0.10 $\\
SBS0909+532   &$ \equiv 0$       &$ 0.20 \pm  0.03 $&                  &                  &$ 1.7 $&$ 0.19 \pm 0.06 $\\
FBQ0951+2635  &$ 0.12 \pm  0.02 $&$ \equiv 0$       &                  &                  &$ 3.2 $&$-0.03 \pm 0.06 $\\
BRI0952-0115  &$ 0.03 \pm  0.02 $&$ \equiv 0$       &                  &                  &$ 5.4 $&$ 0.12 \pm 0.12 $\\
Q0957+561     &$ 0.02 \pm  0.02 $&$ \equiv 0$       &                  &                  &$ 2.4 $&$-0.05 \pm 0.09 $\\
LBQS1009-0252 &$ \equiv 0$       &$ 0.23 \pm  0.02 $&                  &                  &$ 1.3 $&$-0.14 \pm 0.06 $\\
Q1017-207=J03 &$ \equiv 0$       &$ 0.00 \pm  0.02 $&                  &                  &$ 1.6 $&$-0.03 \pm 0.09 $\\
B1030+074     &$ \equiv 0$       &$ 0.02 \pm  0.04 $&                  &                  &$ 4.1 $&$ 0.38 \pm 0.18 $\\
HE1104-1805   &$ \equiv 0$       &$ 0.07 \pm  0.01 $&                  &                  &$ 1.1 $&$-0.14 \pm 0.09 $\\
PG1115+080    &$ 0.00 \pm  0.03 $&$ 0.01 \pm  0.02 $&$ 0.01 \pm  0.02 $&$ \equiv 0$       &$ 3.1 $&$-0.02 \pm 0.09 $\\
Q1208+1011    &$ 0.03 \pm  0.01 $&$ \equiv 0$       &                  &                  &$ 3.3 $&$ 0.07 \pm 0.09 $\\
HST12531-2914 &$ \equiv 0$       &$ 0.17 \pm  0.34 $&$ 0.08 \pm  0.33 $&$ 0.53 \pm  0.41 $&$ 7.6 $&\\
H1413+117     &$ 0.05 \pm  0.01 $&$ 0.07 \pm  0.01 $&$ \equiv 0$       &$ 0.01 \pm  0.01 $&$ 1.2 $&$ 0.22 \pm 0.04 $\\
HST14176+5226 &$ 0.11 \pm  0.14 $&$ \equiv 0$       &$ 0.12 \pm  0.16 $&$ 0.03 \pm  0.17 $&$ 6.1 $&\\
B1422+231     &$ 0.04 \pm  0.01 $&$ \equiv 0$       &$ 0.01 \pm  0.01 $&$ 0.03 \pm  0.01 $&$ 0.5 $&$ 0.35 \pm 0.07 $\\
SBS1520+530   &$ 0.09 \pm  0.01 $&$ \equiv 0$       &                  &                  &$ 2.4 $&$-0.19 \pm 0.06 $\\
B1600+434     &$ \equiv 0$       &$ 0.10 \pm  0.03 $&                  &                  &$ 6.0 $&$ 0.22 \pm 0.09 $\\
PKS1830-211   &$ \equiv 0$       &$ 3.00 \pm  0.13 $&                  &                  &$ 4.1 $&$ 0.57 \pm 0.13 $\\
MG2016+112    &$ \equiv 0$       &$ 0.07 \pm  0.01 $&                  &                  &$ 3.1 $&$-0.01 \pm 0.08 $\\
HE2149-2745   &$ \equiv 0$       &$ 0.01 \pm  0.01 $&                  &                  &$ 1.4 $&$-0.07 \pm 0.06 $\\
Q2237+0305    &$ 0.07 \pm  0.03 $&$ \equiv 0$       &$ 0.18 \pm  0.03 $&$ 0.17 \pm  0.03 $&$ 1.3 $&$ 0.13 \pm 0.04 $\\

%% file: tab4.tex
Q0142-100                        &$ 0.12 \pm  0.01 $&                  &                  \\
B0218+357                        &$ 0.44 \pm  0.14 $&                  &                  \\
MG0414+0534                      &$ 1.05 \pm  0.08 $&$ 0.35 \pm  0.03 $&$ 0.15 \pm  0.01 $\\
SBS0909+532                      &$ 1.17 \pm  0.13 $&                  &                  \\
FBQ0951+2635                     &$ 0.27 \pm  0.02 $&                  &                  \\
BRI0952-0115                     &$ 0.27 \pm  0.01 $&                  &                  \\
Q0957+561                        &$ 0.89 \pm  0.04 $&                  &                  \\
LBQS1009-0252                    &$ 0.31 \pm  0.02 $&                  &                  \\
Q1017-207=J03                    &$ 0.14 \pm  0.01 $&                  &                  \\
B1030+074                        &$ 0.06 \pm  0.01 $&                  &                  \\
HE1104-1805                      &$ 0.29 \pm  0.01 $&                  &                  \\
PG1115+080                       &$ 0.70 \pm  0.05 $&$ 0.17 \pm  0.01 $&$ 0.26 \pm  0.02 $\\
Q1208+1011                       &$ 0.21 \pm  0.01 $&                  &                  \\
H1413+117                        &$ 0.94 \pm  0.03 $&$ 0.66 \pm  0.02 $&$ 0.59 \pm  0.02 $\\
B1422+231                        &$ 1.08 \pm  0.04 $&$ 0.56 \pm  0.02 $&$ 0.03 \pm  0.01 $\\
SBS1520+530                      &$ 0.42 \pm  0.01 $&                  &                  \\
B1600+434                        &$ 0.92 \pm  0.07 $&                  &                  \\
PKS1830-211                      &$ 0.63 \pm  0.07 $&                  &                  \\
MG2016+112                       &$ 1.08 \pm  0.06 $&                  &                  \\
HE2149-2745                      &$ 0.24 \pm  0.01 $&                  &                  \\
Q2237+0305                       &$ 0.66 \pm  0.06 $&$ 0.53 \pm  0.05 $&$ 0.42 \pm  0.04 $\\

%% file: paper.bbl
\begin{thebibliography}{}

\bibitem[]{}Bahcall, J.N., Maoz, D., Schneider, D.P., Yanny, B. \& Doxsey, R., 1992, ApJL, 392, L1
\bibitem[]{}Bernstein, G., Fischer, P., Tyson, J.A. \& Rhee, G., 1997, ApJL, 483, L79
\bibitem[]{}Biggs, A.D., Browne, I.W.A., Helbig, P. \& Koopmans, L.V.E., 1998, astro-ph/9811282
\bibitem[]{}Blanton, M., Turner, E.L. \& Wambsganss, J., 1998, astro-ph/9805359 
\bibitem[]{}Brosch, N. \& Loinger, F., 1991, A\&A, 249, 327
\bibitem[]{}Burstein, D. \& Heiles, C., 1978, ApJ, 225, 40
\bibitem[]{}Cardelli, J.A., Clayton, G.C. \& Mathis, J.S., 1989, ApJ, 345, 245
\bibitem[]{}Carilli, C.L., Rupen, M.P. \& Yanny, B., 1993, ApJL, 412, L59
\bibitem[]{}Chavushyan, V.H., Vlasyuk, V.V., Stepanian, J.A. \& Erastova, L.K., 1997, A\&A, 318, L67 
\bibitem[]{}Chae, K.-H. \& Turnshek, D.A., 1998, astro-ph/9810464
\bibitem[]{}Cheng, Y.-C. \& Krauss, L.M., 1998, astro-ph/9810393
\bibitem[]{}Chiba, M. \& Yoshii, Y., 1998, astro-ph/9808321
\bibitem[]{}Combes, F. \& Wiklind, T., 1997, ApJ, 486, 79
\bibitem[]{}Conner, S.R., Leh\'ar, J. \& Burke, B.F., 1992, ApJL, 387, L61 
\bibitem[]{}Cooray, A.R., 1998, astro-ph/9811448
\bibitem[]{}Corrigan, R.T., Irwin, M.J., Arnaud, J., Fahlman, G.G., Fletcher, J.M., Hewett, P.C., Hewitt, J.N., 
  Le Fevre, O., McClure, R., Pritchet, C.J., Schneider, D.P., Turner, E.L., Webseter, R.L. \& Yee, H.K.C.,
  1991, AJ, 102, 34
\bibitem[]{}Dolan, J.F., Michalitsianos, A.G., Thompson, R.W., Boyd, P.T., Wolinski, K.G., Bless, R.C., Nelson,
     M.J., Percival, J.W., Taylor, M.J., Elliot, J.L. \& Van Citters, G.W., 1995, ApJ, 442, 87 
\bibitem[]{}Draine, B. \& Malhotra, S., 1993, ApJ, 414, 632
\bibitem[]{}Elvis, M., Wilkes, B.J., McDowell, J.C., Green, R.F., Bechtold, J., Willner, S.P., Oey, M.S.,
  Polomski, E. \& Cutri, R., 1994, ApJS, 95, 1
\bibitem[]{}Falco, E.E., Kochanek, C.S. \& Mu\~noz, J.A., 1998, ApJ, 494, 47
\bibitem[]{}Falco, E.E., Leh\'ar, J., Perley, R.A., Wambsganss, J. \& Gorenstein, M.V., 1996, AJ, 112, 897
\bibitem[]{}Fitzpatrick, E.L,. \& Massa, D., 1988, ApJ, 307, 734
\bibitem[]{}Fitzpatrick, E.L, 1998, astro-ph/9809387
\bibitem[]{}Francis, P.J., Hewett, P.C., Foltz, C.B., Chaffee, F.H., Weymann, R.J. \& Morris, S.L., 1991, ApJ, 373, 465
\bibitem[]{}Freedman, W.L., Mould, J.R., Kennicutt, R.C. \& Madore, B.F., 1998,
  in Cosmological Parameters and the Evolution of the Universe, IAU 183, 
  astro-ph/9801080
\bibitem[]{}Frye, B., Welch, W.J. \& Broadhurst, T., 1997, ApJL, 478, L25
\bibitem[]{}Gerin, M., Phillips, T.G., Benford, D.J., Young, K.H., Menten, K.M. \& Frye, B., 1997, ApJL, 488, L31
\bibitem[]{}Gordon, K.D. \& Clayton, G.C., 1998, astro-ph/9802003
\bibitem[]{}Goudfrooij, P. \& de Jong, T. 1995, A\&A, 298, 784
\bibitem[]{}Goudfrooij, P., de Jong, T., Hansen, L. \& Norgaard-Nielsen, H.U., 1994, MNRAS, 271, 833
\bibitem[]{}Hartwick, F.D.A. \& Schade, D., 1990, ARA\&A, 28, 437
\bibitem[]{}Hewett, P.C., Irwin, M.J., Foltz, C.B., Harding, M.E., Corrigan, R.T., 
   Webster, R.L. \& Dinshaw, N., 1994, AJ, 108, 1534
\bibitem[]{}Hodge, P.W. \& Kennicutt, R.C., 1982, 87, 264
\bibitem[]{}Impey, C.D., Falco, E.E., Kochanek, C.S., Leh\'ar, J., McLeod, B., 
               Rix, H.-W., Peng, C. \& Keeton, C.R., 1998, ApJ in press, astro-ph/9803207
\bibitem[]{}Impey, C.D., Foltz, C.B., Petry, C.E., Browne, I.W.A. \& Patnaik, A.R., 1996, ApJL, 462, L53
\bibitem[]{}Impey, C.D. \& Neugebauer, G., 1988, AJ, 95, 307
\bibitem[]{}Iye, M. \& Richter, O.G., 1985, A\&A, 144, 471
\bibitem[]{}Jackson, N., De Bruyn, A.G., Myers, S., Bremer, M.N.,
                    Miley, G.K., Schilizzi, R.T., Browne, I.W.A., Nair, S.,
                    Wilkinson, P.N., Blandford, R.D., Pearson, T.J. \&
                    Readhead, A.C.S., 1995, MNRAS, 274, 25 
\bibitem[]{}Jean, C. \& Surdej, J., 1998, astro-ph/9810218
\bibitem[]{}Jenniskens, P. \& Greenberg, J.M., 1993, A\&A, 274, 439
\bibitem[]{}Keeton, C.R., Kochanek, C.S. \& Falco, E.E., 1998, ApJ 509, 561
\bibitem[]{}King, L.J., Jackson, N., Blandford, R.D., Bremer, M.N., Browne, I.W.A., de Bruyn, A.G., Fassnacht, C.,
  Koopmans, L., Marlow, D. \& Wilkinson, P.N., 1998, MNRAS, 295, 41
\bibitem[]{}Kochanek, C.S., Falco, E.E., Impey, C.D., Leh\'ar, J., McLeod, B.A. \& Rix, H.-W., 1998, astro-ph/9811111 (1998a)
\bibitem[]{}Kochanek, C.S., Falco, E.E., Impey, C.D., Leh\'ar, J., McLeod, B.A., Rix, H.-W., Keeton, C.R.,
  Peng, C.Y. \& Mu\~noz, J.A., 1998, astro-ph/9809371 (1998b)
\bibitem[]{}Kochanek, C.S., Falco, E.E., Schild, R., Dobrzycki, A., Engels, D. \& Hagen, H.-J., 1997, ApJ, 479, 678
\bibitem[]{}Kochanek, C.S., 1996, ApJ, 473, 595
\bibitem[]{}Kormendy, J. \& Stauffer, J., 1987, in Structure \& Dynamics of Elliptical Galaxies, IAU 127, T. De Zeeuw,
 ed., (Kluwer: Dordrecht) 405
\bibitem[]{}Kundi\'c, T., Colley, W.N., Gott, J.R., Malhotra, S., Pen, U.-L., Rhoads, J.E., 
  Stanek, K.Z., Turner, E.L. \& Wambsganss, J., 1995, ApJL, 455, 5
\bibitem[]{}Lawrence, C.R., Neugebauer, G. \& Matthews, K., 1993, AJ, 105, 17 
\bibitem[]{}Leh\'ar, J.,  Falco, E.E., Impey, C.D., Kochanek, C.S., McLeod, B.A., Rix, H.-W. et al.,
   1999, in preparation 
\bibitem[]{}Lovell, J.E.J., Jauncey, D.L., Reynolds, J.E., Wieringa, M.H., King, E.A., Tzioumis, A.K.,
    McCulloch, P.M. \& Edwards, P.G., 1998, astro-ph/9809301
\bibitem[]{}Lovell, J.E.J., Reynolds, J.E., Jauncey, D.L., Backus, P.R., McCulloch, P.M., Sinclair, M.W.,
  Wilson, W.E., Tzioumis, A.K., King, E.A., Gough, R.G., Ellingsen, S.P., Phillips, C.J., Preston, R.A.,
  \& Jones, D.L., 1996, ApJL, 472, L5
\bibitem[]{}Madau, P., Pozzetti, L. \& Dickinson, M., 1998, ApJ, 498, 106
\bibitem[]{}Malhotra, S., Rhoads, J.E. \& Turner, E.L., 1997, MNRAS, 288, 138
\bibitem[]{}Masci, F.J., Webster, R.J. \& Francis, P.J., 1998, MNRAS in press, astro-ph/9808337
\bibitem[]{}Mathis, J.S., 1990, ARA\&A, 28, 37
\bibitem[]{}Mathis, J.S. \& Cardelli, J.A., 1992, ApJ, 398, 610 
\bibitem[]{}Mathur, S. \& Nair, S., 1997, ApJ, 484, 140
\bibitem[]{}McLeod, B.A., Falco, E.E., Impey, C.D., Kochanek, C.S., Leh\'ar, J., Rix, H.-W. et al.,
   1998, in preparation 
\bibitem[]{}McLeod, B.A., Bernstein, G.M., Rieke, M.J. \& Weedman, D.W., 1998, AJ, 115, 1377
\bibitem[]{}Menten, K.M. \& Reid, M.J., 1996, ApJL, 465, L99
\bibitem[]{}Monier, E.M., Turnshek, D.A. \& Lupie, O.L., 1998, ApJ, 496, 177
\bibitem[]{}Moore, C.B. \& Hewitt, J.N., 1997, ApJ, 491, 451
\bibitem[]{}Nadeau, D., Yee, H.K.C., Forrest, W.J., Garnett, J.D., Ninkov, Z. \& Pipher,
  J.L., 1991, ApJ, 376, 430
\bibitem[]{}Neugebauer, G., Soifer, B.T., Mathews, K. \& Elias, J.H., 1989, AJ, 97, 957
\bibitem[]{}Ostensen, R., Remy, M., Lindblad, P.O., Refsdal, S., Stabell, R., Surdej, J., Barthel, P.D., et al.,
  1996, A\&AS, 126, 393
\bibitem[]{}Ostensen, R., Refsdal, S., Stabell, R., Teuber, J., Emanuelsen, P. I., Festin, L., Florentin-Nielsen, R., et al.,
  1996, A\&A, 309, 590
\bibitem[]{}Patnaik, A.R., Browne, I.W.A., Walsh, D., Chaffee, F.H. \& Foltz, C.B., 1992, MNRAS, 259, 1p
\bibitem[]{}Patnaik, A.R., Porcas, R.W. \& Browne, I.W.A., 1995, MNRAS, 274, 5p 
\bibitem[]{}Perlmutter, S., Gabi, S., Goldhaber, G., Goobar, A., Groom, D.E., Hook, I.M., 
  Kim, A.G., et al., 1997, ApJ, 483, 565
\bibitem[]{}Ratnatunga, K.U., Ostrander, E.J., Griffiths, R.E. \& Im, M., 1995, ApJL, 453, L5 
\bibitem[]{}Remy, M., Surdej, J., Smette, A. \& Claeskens, J.-F., 1993, A\&A, 278, L19 
\bibitem[]{}Riess, A. G. 1998, private communication
\bibitem[]{}Riess, A.G., Filippenko, A.V., Challis, P., Clocchiatti, A., Diercks, A., 
   Garnavich, P.M., Gilliland, R.L., et al., 1998, AJ, 116, 1009
\bibitem[]{}Riess, A.G., Press, W. \& Kirshner, R.P., 1996, ApJ, 473, 588
\bibitem[]{}Rifatto, A., 1990, in Dusty Objects in the Universe, E. Bussoletti \& A.A. Vittone,
  eds., (Kluwer: Dordrecht) 277
\bibitem[]{} Rix, H.-W., Schneider, D.P. \& Bahcall, J.N., 1992, AJ, 104, 959 
\bibitem[]{}Rouleau, F., Henning, T. \& Stognienko, R., 1997, A\&A, 322, 633
\bibitem[]{}Saust, A. B. 1994, A\&AS, 103, 33
\bibitem[]{}Savage, B.D. \& Mathis, J.S., 1979, ARA\&A 17, 73
\bibitem[]{}Savage, B.D., Bohlin, R.C., Drake, J.F., Budich, W., 1977, ApJ, 216, 291 
\bibitem[]{}Schechter, P.L., Gregg, M.D., Becker, R.H., Helfand, D.J. \& White, R.L., 1998, AJ, 115, 1371
\bibitem[]{}Schechter, P.L., 1998, private communication
\bibitem[]{}Schlegel, D.J., Finkbeiner, D.P. \& Davis, M., 1998, ApJ, 500, 525
\bibitem[]{}Schneider, D.P., Gunn, J.E., Turner, E.L., Lawrence, C.R., Hewitt, J.N., Schmidt, M.,
   \& Burke, B.F., 1986, AJ, 91, 991
\bibitem[]{}Schneider, D.P., Lawrence, C.R., Schmidt, M., Gunn, J.E., Turner, E.L., Burke, B.F.,
    \&  Dhawan, V., 1985, ApJ, 294, 66 
\bibitem[]{}Schneider, P., Ehlers, J. \& Falco, E.E., 1992, Gravitational Lenses, (Springer: Berlin)
\bibitem[]{}Schweizer, F.,, 1987, in Structure \& Dynamics of Elliptical Galaxies, IAU 127, T. De Zeeuw,
 ed., (Kluwer: Dordrecht) 109
\bibitem[]{}Turnshek, D.A., Lupie, O.L., Rao, S.M., Espey, B.R. \& Sirola, C.J., 1997, ApJ, 485, 100 
\bibitem[]{}Ulrich, M.-H., Maraschi, L. \& Megan, C., 1997, ARA\&A, 35, 445
\bibitem[]{}van Dokkum, P.G. \& Franx, M., 1995, AJ, 110, 2027
\bibitem[]{}Wagner, S., 1998, private communication
\bibitem[]{}Walterbos, R., 1986, PhD thesis, Leiden University.
\bibitem[]{}Warren-Smith, R.F. \& Berry, D.S., 1983, MNRAS, 205, 889
\bibitem[]{}Webster, R.L., Francis, P.J., Peterson, B.A., Drinkwater, M.J. \& Masci, F.J., 1995, Nature, 375, 469
\bibitem[]{}Wiklind, T. \& Combes, F., 1995, A\&A, 299, 382
\bibitem[]{}Wiklind, T. \& Combes, F., 1996, Nature, 379, 139
\bibitem[]{}Wiklind, T. \& Combes, F., 1998, ApJ, 500, 129
\bibitem[]{}Wills, B.J., Wills, D., Evans, N.J., Natta, A., Thompson, K.L., Breger, M. 
  \& Sitko, M.L., 1992, ApJ, 400,96
\bibitem[]{}Wisotzki, L., Kohler, T., Lopez, S. \& Reimers, D., 1996, A\&A, 315, L405
\bibitem[]{}Witt, A., Thronson, H. \& Capuano, J., 1992, ApJ, 393, 611
\bibitem[]{}Xanthopoulos, E., Browne, I.W.A., King, L.J., Koopmans, L.V.E., Jackson, N.J., Marlow, D.R.,
   Patnaik, A.R., Porcas, R.W. \& Wilkinson, P.N., 1998, astro-ph/9802014 
\bibitem[]{}Yee, H.K.C. \& Ellingson, E., 1994, AJ, 107, 28

\end{thebibliography}
